%% file: main.tex
\documentclass[11pt,letterpaper]{article}

\usepackage{flushend}
\usepackage{dsfont}
\usepackage[margin=1in]{geometry}

\usepackage{rotating}
\usepackage{graphicx}
\usepackage{amssymb}
\usepackage{amsmath}
\usepackage{epstopdf}
\usepackage[hyphens]{url}
\usepackage[small,compact]{titlesec}

\usepackage[pdftex]{hyperref}
\hypersetup{colorlinks=false, pdfborder= 0 0 0}

\usepackage{color}
\usepackage{soul}
\usepackage{times}
\usepackage{setspace}
\usepackage{subfigure}
\usepackage{hyperref}
\usepackage{booktabs}
\hypersetup{colorlinks=false, pdfborder= 0 0 0}

\usepackage{bbm}

\usepackage[draft,inline,nomargin]{fixme}

\usepackage{xspace}

\newcommand{\sys}{ScreenAvoider\xspace}

\usepackage{authblk}  

\title{\Large \bf ScreenAvoider: Protecting Computer Screens from Ubiquitous Cameras\thanks{Mohammed Korayem and Robert Templeman contributed equally.}  }

\newcommand{\xhdr}[1]{ \vspace{1pt} \noindent {\textbf{\textit{#1}}}}
\newcommand{\leaveout}[1]{}

\author{Mohammed Korayem,$^{\dag}$ Robert Templeman,$^{\dag\ddag}$ Dennis Chen,$^{\flat}$ David Crandall,$^{\dag}$ Apu Kapadia$^{\dag}$\\
$^{\dag}$School of Informatics and Computing\\
Indiana University Bloomington\\
Bloomington, IN, USA\\
\{retemple, mkorayem, djcran, kapadia\}@indiana.edu\\
\vspace{2mm}

$^{\ddag}$Naval Surface Warfare Center\\
Crane, IN, USA\\
robert.templeman@navy.mil\\
\vspace{2mm}

$^{\flat}$Olin College\\
Needham, MA, USA\\
Dennis.Chen@students.olin.edu\\
\vspace{-3mm}
}

\date{}

\begin{document}
\maketitle
\thispagestyle{empty}
\input{abstract}

\input{intro}

\input{overview}

\input{eval}
\input{discussion}

\input{related}

\input{conclusion}

\section*{Acknowledgment}
This material is based upon work supported by the National Science Foundation under grants CNS-1408730 and IIS-1253549, and a Google Faculty Research Award. Any opinions, findings, and conclusions or recommendations expressed in this material are those of the author(s) and do not necessarily reflect the views of the sponsors.

\bibliographystyle{IEEEtran}
\bibliography{IEEEabrv,refs}
\end{document}

%% file: abstract.tex
\begin{abstract}

We live and work in environments that are inundated with cameras embedded in devices such as phones, tablets, laptops, and monitors. Newer wearable devices like Google Glass, Narrative Clip, and Autographer offer the ability to quietly log our lives with cameras from a `first person' perspective. While capturing several meaningful and interesting moments, a significant number of images captured by these wearable cameras can contain computer screens. Given the potentially sensitive information that is visible on our  displays, there is a need to guard computer screens from undesired photography. People need protection against photography of their screens, whether by other people's cameras or their own cameras.  

We present \sys, a framework that controls the collection and disclosure of images with computer screens and their sensitive content. \sys can detect images with computer screens with high accuracy and can even go so far as to discriminate amongst screen content. We also introduce a ScreenTag system that aids in the identification of screen content, flagging images with highly sensitive content such as messaging applications or email webpages. We evaluate our concept on realistic lifelogging datasets, showing that \sys provides a practical and useful solution that can help users manage their privacy.

\end{abstract}

%% file: intro.tex

\section{Introduction}
\label{sec:intro}

\begin{figure}[t]
\begin{center}
\includegraphics[width=5in]{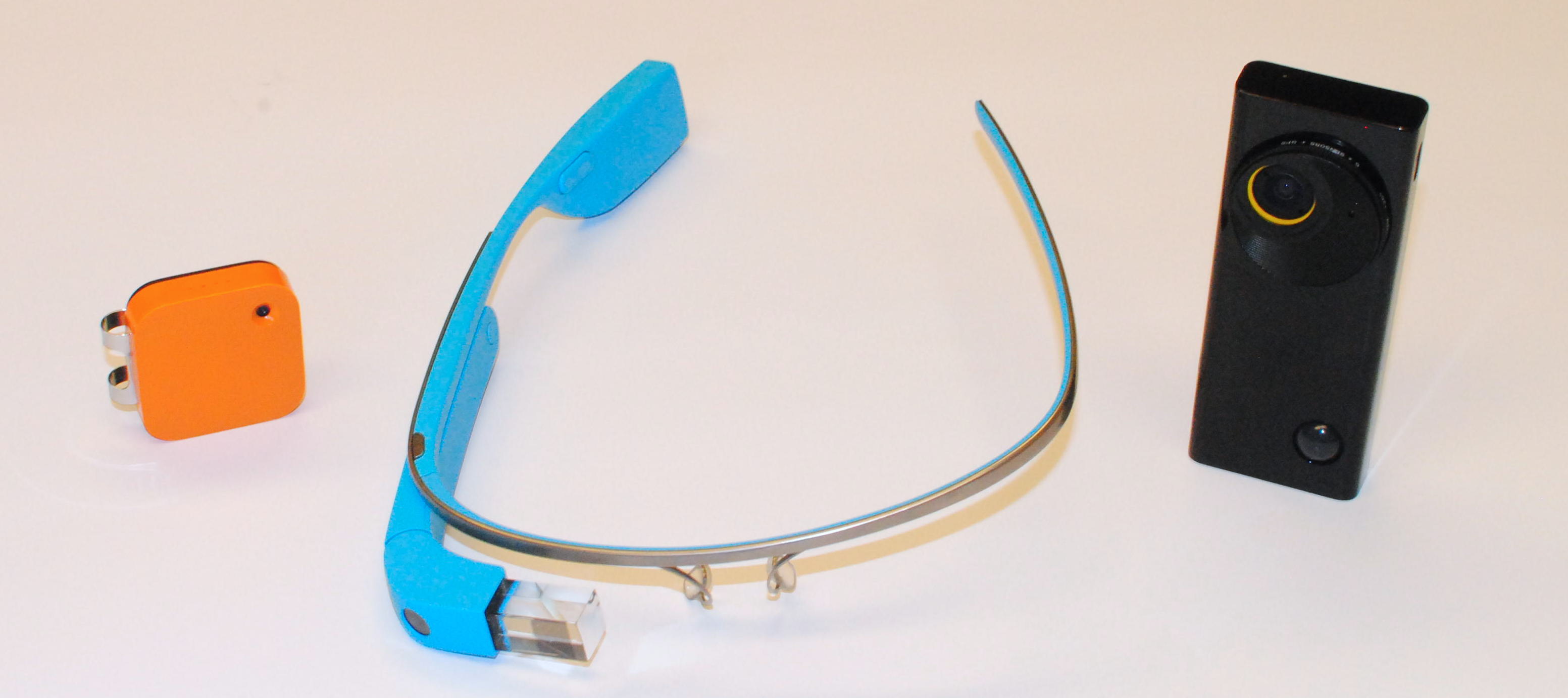} 
\end{center}
\caption{Wearable cameras: From left, Narrative Clip, Google Glass, and Autographer.}
\label{fig:cameras}
\end{figure}

Cameras are pervasive and their numbers continue to grow. In addition
to surveillance cameras installed on streets and in businesses, most people
now own and carry around multiple  cameras,
since modern laptops, smartphones, tablets, monitors, gaming systems, televisions, and home automation systems are now equipped with cameras by default.
Meanwhile, wearable cameras like Google Glass~\cite{glass}, the Narrative
Clip~\cite{narrative}, and Autographer~\cite{autographer} 
have recently come on the market, allowing people to record their
whole lives from a `first-person' perspective
(Figure~\ref{fig:cameras}).  These wearable devices enable useful
applications like allowing users to take visual diaries of their lives
(a concept known as `lifelogging'), for instance to help improve their
personal security, to treat memory loss and forms of
dementia~\cite{sensecam}, or just for fun.

These wearable cameras can collect \emph{thousands} images
every day, many of which  may capture private
activities (like using the restroom) or information (like catching private documents)~\cite{hoyle2014privacy}.
Many of these devices communicate
with cloud-based applications, so that images are automatically shared
with the cloud provider, and software features make it as easy to
share images as it is to collect them. These features raise obvious
privacy concerns, as research shows that users themselves often
mistakenly disclose information electronically (through
`misclosures')~\cite{Caine2009}. This problem is exacerbated with
large, unwieldy collections of lifelogging images, any of which may contain risks to privacy. Moreover one must
trust the security of the cloud where many images are stored and used,
with the recent case of celebrity photos stolen from hacked iCloud
accounts showing that cloud photo storage is not always
secure~\cite{clouddistrust}. Ideally, sensitive images should be
kept off the cloud and possibly deleted completely.

\begin{figure*}[t]
\begin{center}
\fbox{\includegraphics[width=1.1in,height=.825in]{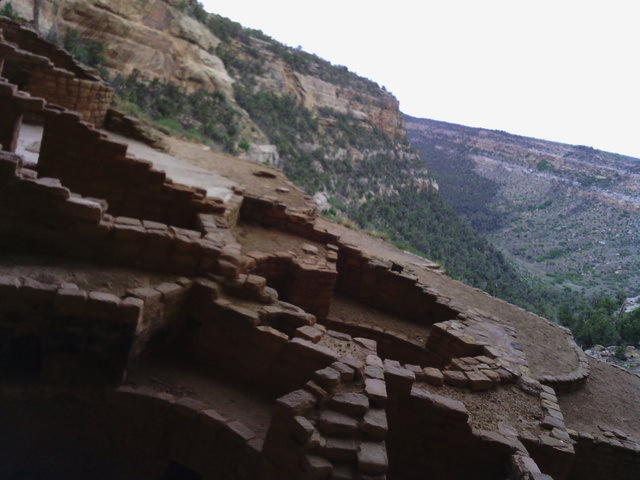}
\includegraphics[width=1.1in,height=.825in]{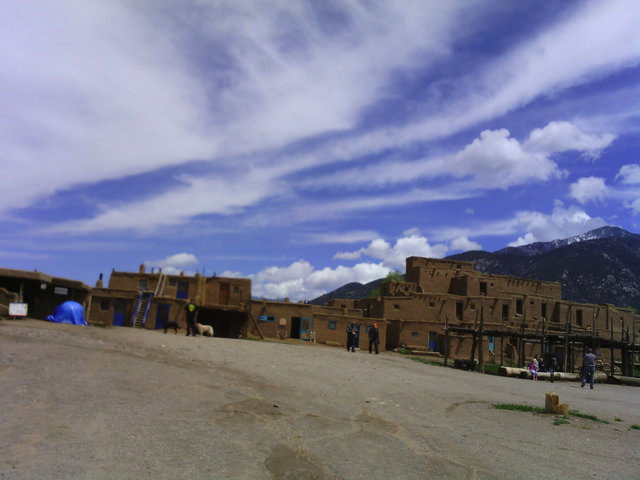}
\includegraphics[width=1.1in,height=.825in]{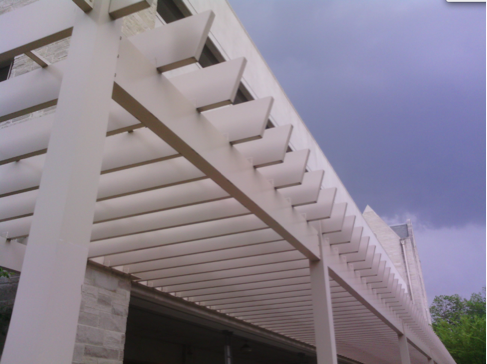}
\includegraphics[width=1.1in,height=.825in]{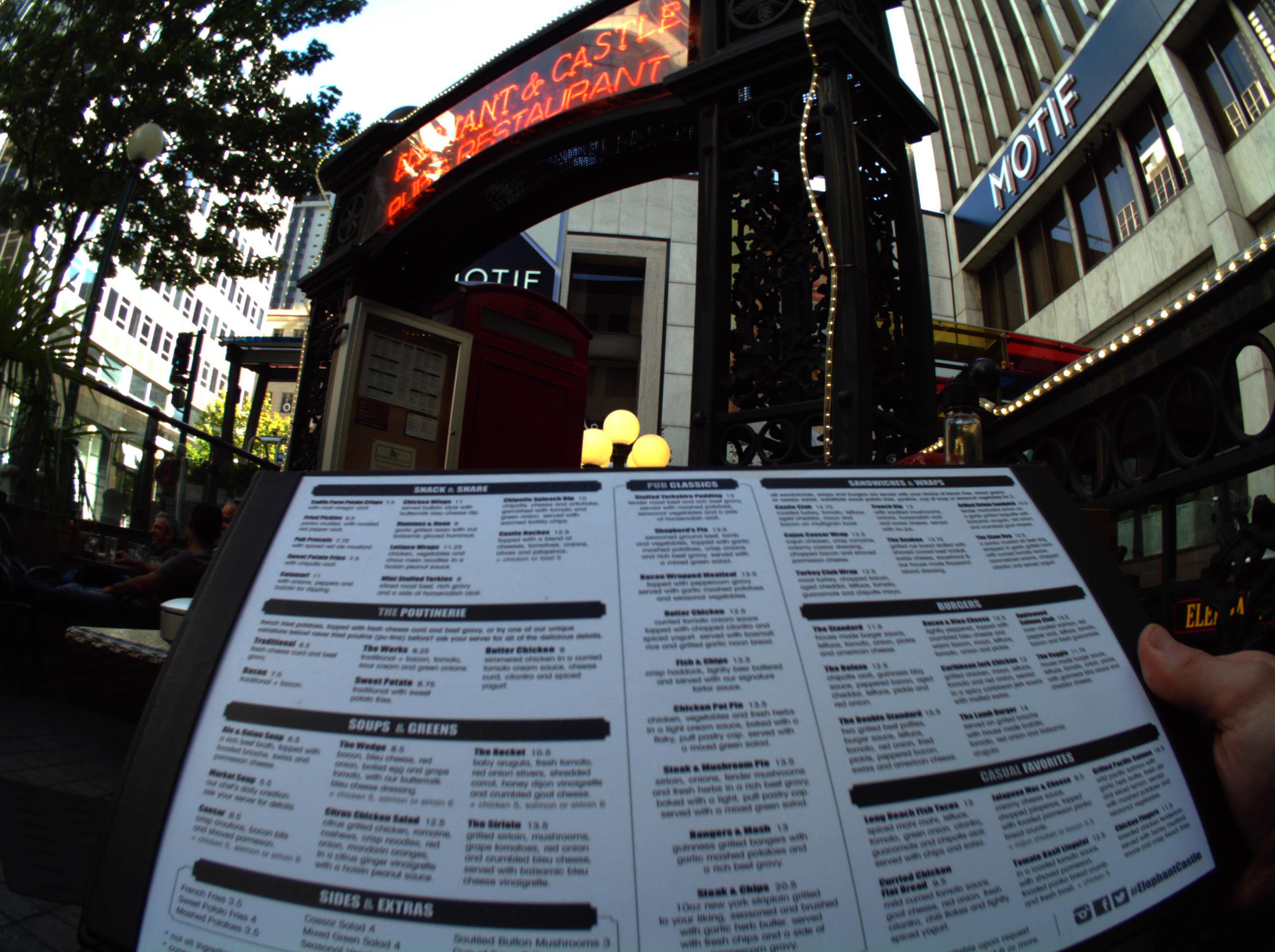}}\\
interesting\\ 
\vspace{12pt}
\fbox{\includegraphics[width=1.1in,height=.825in]{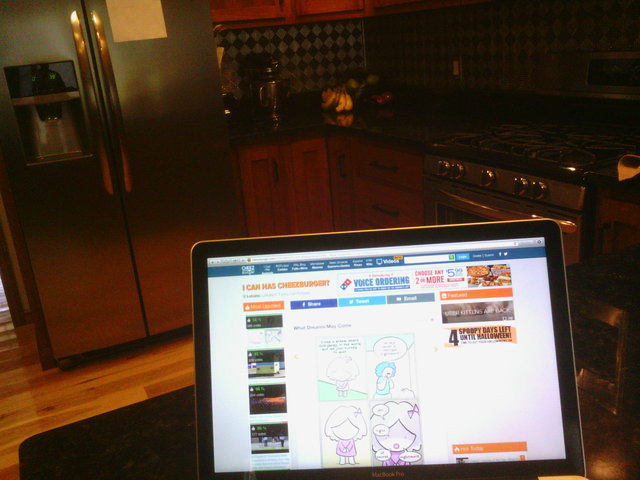}
\includegraphics[width=1.1in,height=.825in]{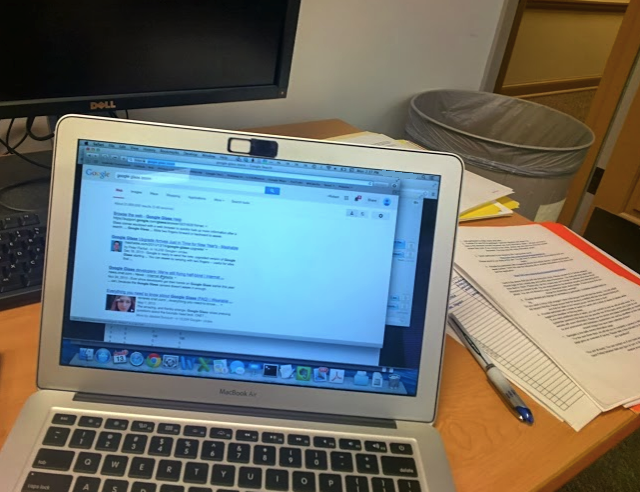}}
\fbox{\includegraphics[width=1.1in,height=.825in]{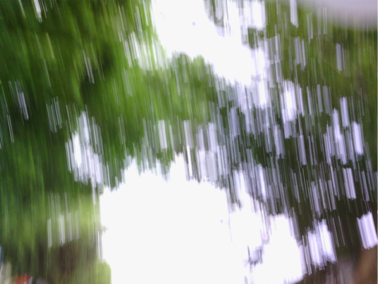}
\includegraphics[width=1.1in,height=.825in]{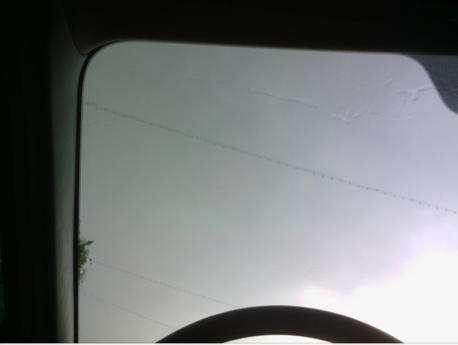}}\\
sensitive \hspace{2in} useless\\
\end{center}
\caption{Some sample lifelogging images, showing that first-person
  cameras capture a mix of photos including  interesting images (top), sensitive images of
  monitors (bottom-left panel), and useless images (bottom-right
  panel).}
\label{fig:lifelog-variety}
\end{figure*}

We thus need techniques for helping users control how images from
wearable cameras are collected and shared. Some very recent research
has considered this problem. For example, Klemperer et
al.~\cite{klemperer2012} suggest an access control system based on
image tags that are assigned manually by users. A major difficulty
with this approach is that manually reviewing images from lifelogging
cameras is prohibitively time-consuming, given that these devices capture
images several times per minute, easily collecting thousands of images
per day.  Raval et al. propose MarkIt~\cite{markit}, where users can make annotations
in private areas of a scene (like drawing a box around sensitive
information on a whiteboard) which are recognized by the lifelogging
camera and blurred or obscured.  Roesner et al. propose a
`World-Driven Access Control' (WDAC) framework that relies on
recognizing policy `passports'~\cite{wdac} that are embedded in the
physical world, like barcodes affixed to private objects.  But the
performance of both of these systems is limited by how well the physical world is annotated, and only certain
types of private information can be annotated this way.  

Templeman et al.~\cite{templeman-abac} propose addressing this
limitation through an attribute-based access control (ABAC) framework
that would use computer vision techniques to detect visual attributes of a
scene, allowing users to create policies based on the presence of these
attributes. They present one implementation called
PlaceAvoider~\cite{placeavoider14ndss} that recognizes room scenes with the goal of identifying photos taken in sensitive spaces, e.g., 
allowing users to block images taken in bathrooms and bedrooms. However, they consider only this single location attribute.
Hoyle et al.~\cite{hoyle2014privacy} conducted a study of lifelogging
users and confirmed that location is sometimes an important indicator
of image privacy, but found that \emph{other attributes like the
  presence of specific objects, especially computer monitors, are of
  much more concern to lifeloggers}. The finding that computer
displays are a common concern is perhaps not surprising, given that
the average American adult spends more than five hours a day in front
of a digital device~\cite{screentime}. 

In this paper we address this
specific problem of detecting and classifying images with computer
displays, to help people protect sensitive information that is routinely
displayed on their screens (like e-mails, instant messages, financial information,
personnel records, etc.). We call this framework ``\sys.''
To help understand the features of \sys, we first provide a motivating example:

\vspace{8pt}
\begin{quotation}
\noindent
\emph{Mary wears an Autographer lifelogging device to record her
  life. She uses a cloud-based lifelog archival service to curate her
  images. This service allows her to define policies based on where
  images were taken.  She has a (PlaceAvoider) policy that marks photos from her
  office as \emph{private} and photos taken in public places as
  \emph{public}. Additionally she likes to keep her office images off the cloud. Today Mary decides to take her laptop to a local
  caf\'{e} for a working lunch.}
\end{quotation}
\vspace{8pt}

Mary's policy reflects that she views private information
(e.g. student grades) and conducts other private business in her office.

\vspace{8pt}
\begin{quotation}
\emph{As Mary begins working at the caf\'{e}, she remembers that her
  lifelog service supports detecting images of monitors through a
  `ScreenAvoider' policy. These policies allow her to define sharing
  preferences based on the presence of computer monitors in her
  images. She quickly enables a policy that prevents sharing images
  containing a computer screen. When she gets home in the evening and
  reviews her lifelogs, she realizes that there are many images of her
  playing Minecraft that she wants to share with her friends.  She
  revises her ScreenAvoider policy to prevent images with her email or
  instant messenger applications from being uploaded to the cloud or
  shared with her friends by a cloud service.}
\end{quotation}

\vspace{8pt}

As this example illustrates, simply detecting the presence of monitors
may be useful to some users, but many will want to define policies
based on finer-grained attributes like \textit{what} is displayed on
the monitor. Indeed, Hoyle et al. also found that many people wanted to share at least some images having monitors  with some social contacts. 
Blocking all monitors in one's lifelogs would mean effectively
erasing the five or more hours of their day that they spend interacting
with the virtual world.

With \sys, therefore, we aim to provide users with a way to specify privacy policies based on  a) whether images contain computer monitors, and b) which applications are displayed on the captured screen.

\begin{figure}[t]
\begin{center}
\fbox{\includegraphics[width=1.65in]{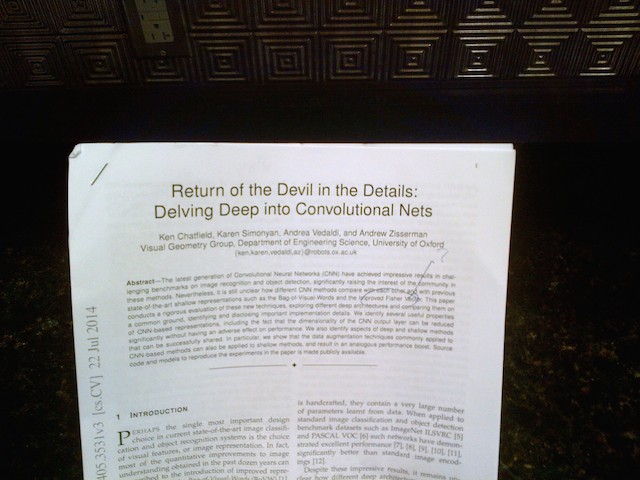}
\includegraphics[width=1.65in]{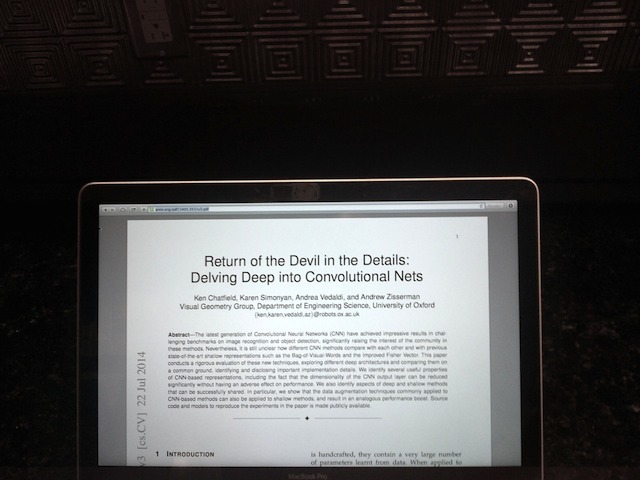}}
\hspace{1mm}
\fbox{\includegraphics[width=1.65in]{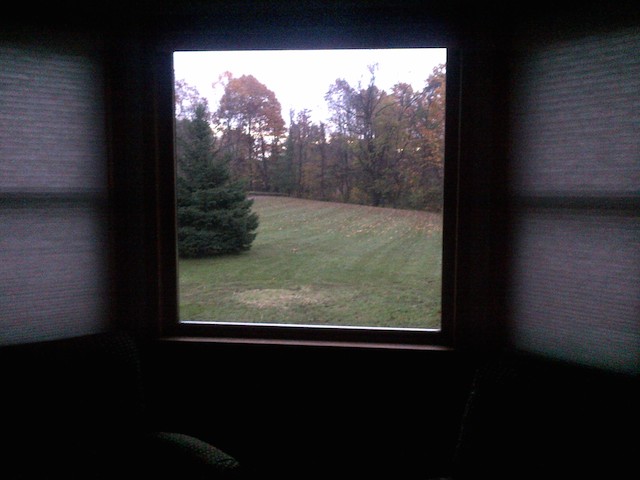}
\includegraphics[width=1.65in]{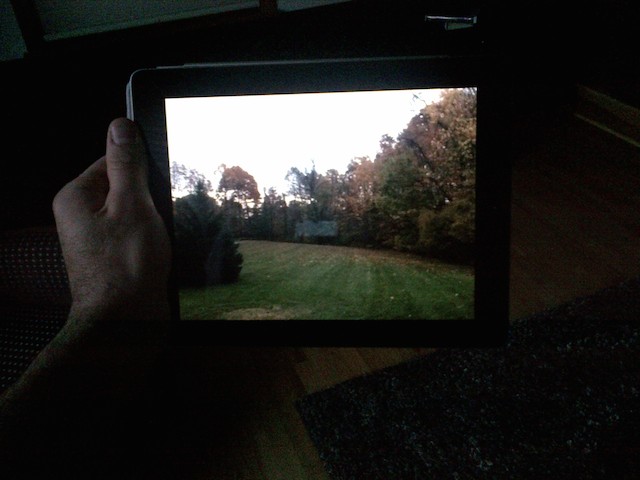}}
\end{center}
\caption{Examples of particularly difficult images for \sys to classify. Each
  row shows two nearly-identical images, one of the real world
  and another displayed by a screen.}
\label{fig:lifelog-examples}
\end{figure}

\xhdr{Research Challenges.} Our work  addresses significant
challenges to make the \sys system work. Detecting monitors and recognizing
their content is a challenging computer vision problem, especially
given that lifelogging images are usually poorly composed (often capturing
portions of monitors at unusual orientations with poor focus and motion blur).
Moreover the content of monitors is so dynamic that it is difficult to define
reliable and distinctive image features, besides very generic properties
like rectangular shape.
Detecting monitors is sometimes difficult even for a human, 
as illustrated in Figure~\ref{fig:lifelog-examples}, since modern monitors can
render photo-realistic scenes that are  hard to distinguish from reality.

However, computer vision techniques have improved dramatically very
recently, due to the emergence of new machine learning techniques
based on \textit{deep learning}. While machine learning has been used
in vision for over a decade, state-of-the-art approaches have
typically used manually-created image features from which classifiers
were learned.  Deep learning is a new paradigm where the image
features are learned with the image classifiers simultaneously,
typically using a Convolutional Neural Network
(CNN)~\cite{krizhevsky2012imagenet} trained on large collections of
images with huge amounts of computation made practical by high-end
Graphics Processing Units (GPUs). These techniques have significantly surpassed a
number of standard benchmarks in other recognition problems, causing
excitement that deep learning may be a large step forward in vision
technology. 

In this paper we present a \sys framework to control pictures that are
taken of our monitors, using deep learning to build models of monitor
images at the granularity of applications. To our knowledge, we are
the first to attempt monitor detection and content recognition, as
well as the first to apply deep learning to lifelogged images. 
Given the difficulty of this problem, we also study an easier variant
of the problem where a custom computer application called ScreenTag
displays machine-readable information on the monitor itself.
This approach is 
in the spirit of the MarkIt~\cite{markit} and WDAC~\cite{wdac}, 
but is updated dynamically and automatically
 based on the current content and sensitivity properties of what is being displayed on the screen. As with WDAC and MarkIt, such policies can be used to control both screen-owners' cameras as well as those carried by other people. However in the case where bystanders' monitors are captured by other lifeloggers, one must rely on the camera owners for filtering out such images. Hoyle et al. found that camera owners have a sense of ``propriety'' where they are unwilling to share images that may violate bystanders' privacy. Their findings indicate that lifeloggers may be willing to use `propriety policies' (e.g., ``I am willing to discard 20\% of my images if they violate other people's privacy'').
 \vspace{2mm}
 
\xhdr{Our Contributions.}
Our specific contributions are:
\begin{enumerate}
   \item \textbf{Presenting \sys}, a framework that can detect lifelogging images (which are often blurry and poorly composed) with computer screens with high accuracy, and even discriminate amongst running applications;
   
   \item \textbf{Introducing ScreenTag}, a service that dynamically creates a recognizable visual element in order to aid \sys;
   
  \item \textbf{Implementing and evaluating \sys} using state-of-the-art deep learning techniques from computer vision, tested on lifelogging images collected from multiple sources to demonstrate the feasibility and limitations of such a system. 
\end{enumerate}

The remainder of this paper describes our contributions in detail. Section~\ref{sec:overview} describes our architecture, constraints,
and concept of operation, while Section~\ref{sec:eval} reports our evaluation
on several first-person datasets.  We discuss the implications of our
results in Section~\ref{sec:disc} before surveying related work in
Section~\ref{sec:related} and concluding in
Section~\ref{sec:conclusion}.

%% file: overview.tex

\section{Our Approach}
\label{sec:overview}

We now explain the \sys framework for detecting images with monitors
and specific types of on-screen content in detail.
We begin by outlining our privacy goals and the adversary model.

\subsection{Privacy goals and adversary model} 

Unlike with imagery taken from point-and-shoot cameras, where the
photographer deliberately composes the scene, with wearable cameras
the lifeloggers play the role of a `curator' who must sift through and
identify the interesting photos that are worth sharing and those that
should be withheld or deleted.  Our high-level objective for \sys is
to enhance a curatorial tool, e.g., one based on ABAC as proposed by
Templeman et al., that reduces the workload for users in finding their
private photos.  We specifically target monitors because the
lifelogging study by Hoyle et al. found that computer monitors were
the single most frequent reason people chose not to share their
photos: of the 10\% of images that the users did not share, 30\%
contained monitors~\cite{hoyle2014privacy}.  Computer monitors
occurred in 30\% of the images (based on a random sample), and of these 87\% were actually
shared.  
  Our main objective of \sys is to address this
privacy concern by automatically identifying images with monitors, as
well as to identify the content on the monitors, since some
applications typically include private information while others show
information that may be benign or even desirable to share.
(As users of lifelogging devices ourselves, we informally confirmed
that computer monitors represent the most frequent potential privacy
leaks for us as well.)

 Our problem reduces to an information retrieval task where images
 with monitors, or images with monitors that display \emph{specific
   applications}, are identified and handled appropriately. The
 application of lifelogging also offers some leeway in terms of
 precision. Whereas it may be important to have high recall rates so
 that all sensitive monitors are identified, having moderate precision
 rates (i.e. relatively frequent false positives) may be acceptable
 (since with thousands of images being randomly captured per day, it
 may not matter much if some are censored unnecessarily).  Of course,
 the exact best trade-off between precision and recall is likely to be
 application-specific, so we do not make any judgments on what this
 tradeoff may be and present complete Precision-Recall curves in
 Section \ref{sec:eval}. In practice, users could specify how
 conservative they want the detection to be, while being cognizant of
 the number of images that may be falsely blocked.
 
 Of course, even at a relatively low precision, we cannot hope for
 perfect recall. Like Raval et al.~\cite{markit}, we do not believe this is a fatal problem:
 while it may be impossible to prevent the leakage of certain `smoking
 gun' types of information, there are several other types of
 situations where privacy improves as more and more (e.g.,
 embarrassing) content is removed. Thus while \sys may leak private
 information through false negatives, we assume the overall impact of
 preventing the leakage of most sensitive images provides a clear,
 overall benefit to users.
 
In an application, \sys could be used as an
additional component to detect sensitive images either a) at the OS
level to control what types of images are shared with 
untrusted applications or uploaded to an untrusted cloud service, b)
in a cloud service that the user trusts for managing access (by other
users) to his/her lifelogging photo albums, or c) as a mechanism that
is used directly by a sensing device to control collection. In the
first category, Jana et al's work with the Darkly
system~\cite{scannerdarkly} and recognizers~\cite{recognizers} address
the access control problem with a general solution relying on a
limited OpenCV API~\cite{opencv} (which does not support the detection
of monitors) and can leverage our work.

In the following subsections we first describe our system architecture followed by overviews of the screen detection approach and the screen content classifier.

\subsection{System architecture}

\begin{figure}[t]
\begin{center}
{\includegraphics[width=3.5in]{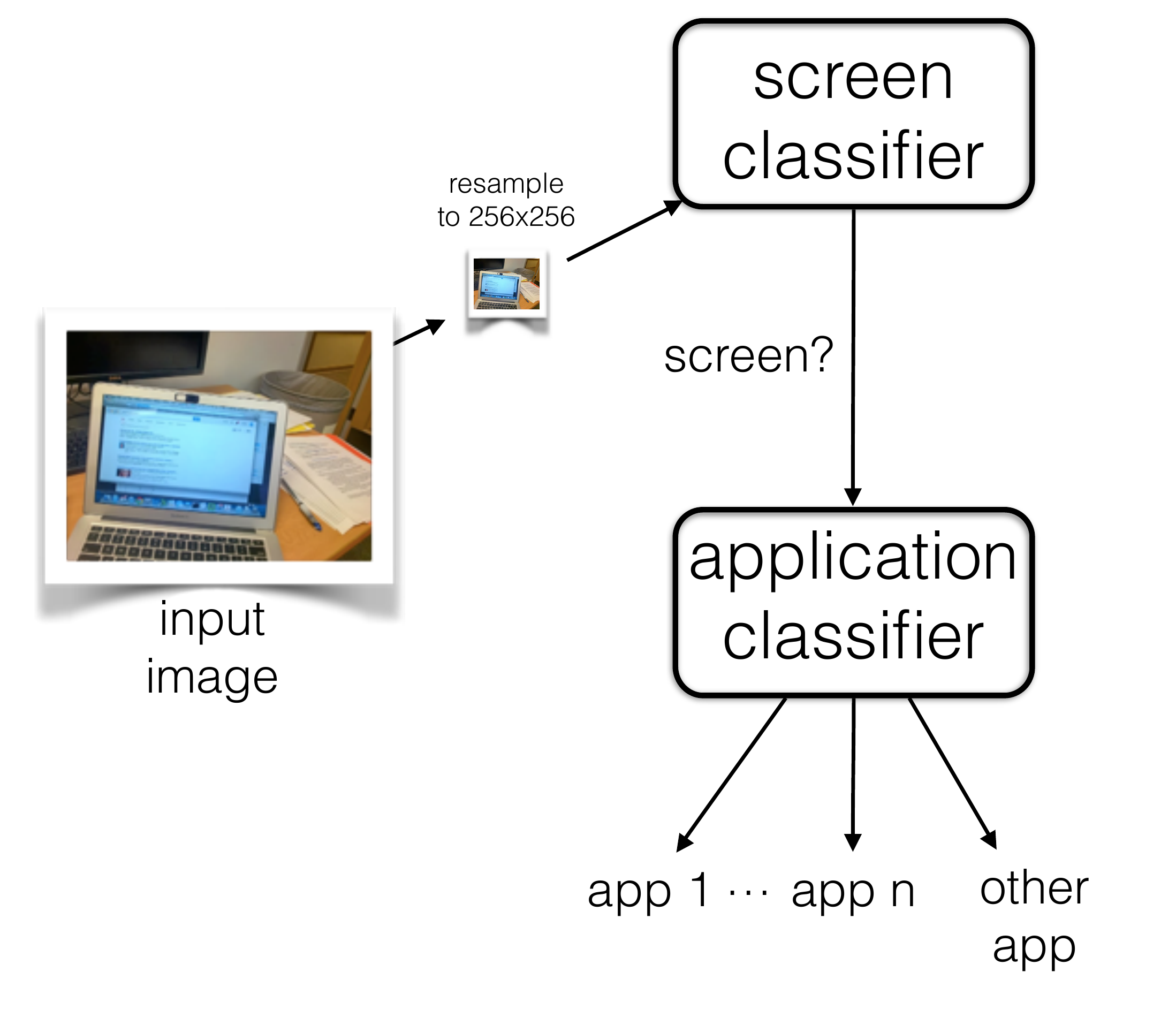}}
\end{center}
\caption{The \sys hierarchical classifier. Native images are downsized for the Caffe CNN framework. While this depiction shows two classification levels, in Subsection \ref{subsec:multiway} we also present a single classifier that includes applications and a class without screens.}
\label{fig:arch}
\end{figure}

Current lifelogging platforms including Autographer and Narrative
Clip, as well as more general-purpose devices like Google Glass that
can run lifelogging applications, offer cloud-based services for
storing and managing of images. Our \sys system permits the
organization of images by their content using a hierarchical
classifier, as illustrated in Figure~\ref{fig:arch}. When presented
with an image, the system uses a classifier trained through machine
learning to first determine whether a screen is visible. If a screen
is detected, the image is passed to a multi-way classifier that
attempts to infer whether any applications of interest are visible.
Because this is a difficult classification problem to perform using
visual features alone, especially when only a portion of the screen is
visible, we have also explored a technique that eases the problem
through a custom application running on the computer itself. This
ScreenTag system, which is complementary to \sys, dynamically creates and renders a machine-readable
visual code overlaid on the computer's display, that contains information
about which applications are running on the system. This way, lifelogging
photos taken of the monitor include a ``watermark'' that is easier
for the lifelogging system to detect and interpret.

\subsection{Detecting computer screens and monitors in images}

Detecting computer screens in images is a specific application of the
general problem of \textit{object category detection} in computer
vision, where the goal is to recognize broad categories of objects
whose visual appearance may vary dramatically from one object to the
next (like cars, airplanes, pedestrians, etc.). Even the same instance of an object
can appear very different from one image to the next, due to
variations in lighting, camera angle, lens zoom, etc. The key
challenge in object category recognition is how to build models that
are invariant to this visual variation that does not relate to the
object's identity, while being sensitive to features that differentiate
monitors from other similar objects (e.g.\ picture frames, windows,
hardcopy print-outs, etc.).

To separate these important visual characteristics from the ones that
should not matter, most work in category recognition takes a machine
learning approach. Low-level visual features are extracted from the
raw pixels of an image, typically corresponding to properties of
color, texture, and shape, and are represented as high-dimensional
vectors in some feature space. Then a discriminative machine learning
algorithm like Support Vector Machines (SVMs) or Random
Forests is given these extracted feature vectors for a set
of images with known ground-truth labels (e.g. monitor and
non-monitors), and the algorithm attempts to learn a decision boundary
between the two classes in the feature space. Given a new image at
classification time, the same features are extracted and the learned
classifier is used to estimate its unknown label.  (We summarize key
related work  in more detail in
Section~\ref{sec:related}).

We applied this traditional category recognition approach to
detecting monitors in lifelogging images, using a battery of
state-of-the-art image features. These included simple image-level features
like color histograms, more advanced scene layout
features like GIST~\cite{gist} and Local Binary Pattern
histograms (which primarily capture global texture), and features that cue on local image regions including
vector-quantized Histograms of Oriented Gradients (HOG)~\cite{dalal05hog} and
SIFT~\cite{sift} features~\cite{placeavoider14ndss}. We then learned image
classifiers with SVMs and thousands of annotated lifelogging images,
and obtained promising preliminary results.

However, during just these few months of preliminary work, a new and
potentially breakthrough technique emerged that has since far surpassed
numerous long-standing benchmarks across a range of computer vision
problems.  Krizhevsky et al.~\cite{krizhevsky2012imagenet} first
reported results on the 2012 ImageNet challenge~\cite{ILSVRCarxiv14}
dataset (which is perhaps the premier object category detection
competition) that significantly cut the recognition error rate using a
technique based on Convolutional Neural Networks. The key idea behind
this approach is that instead of first designing low-level features by
hand and then running a machine learning algorithm, a single unified
algorithm should learn both the low-level features and the
high-classifier simultaneously. 

Krizhevsky et al. showed that this `deep learning' could be accomplished
efficiently using a neural network trained using
backpropagation, very similar to classic techniques
that have been known for many years~\cite{lecun98}. However, they used
more layers (typically seven or more, compared to more traditional
values like three), and vastly more training data (tens to
hundreds of millions of images). Training networks of this size
requires massive amounts of computation, but modern Graphical
Processing Units (GPUs) are well-suited for these calculations since they
primarily involve simple linear algebra operations (e.g. dot
products).

 Here we apply Convolutional Neural Networks to our problem of screen
 detection in lifelogging images.  To our knowledge, no other work has
 studied CNNs with this type of data.  Unfortunately, because widespread
 use of CNNs is so new, not much is known about why these models work
 so well on some problems but not on others. One critical factor is
 that because the networks are so deep and thus have so many
 parameters, they need a very large number of training images to work correctly
 (and otherwise they ``overfit'' to a specific training set instead of
 learning general properties about it).  

A key challenge for applying this approach to lifelogging data is thus the lack of
labeled large-scale training data; even though lifelogging devices
capture several thousand photos per day, actually collecting and
annotating millions of images would be prohibitively expensive. We
tried several techniques to counter this problem, as described in more
detail in Section~\ref{sec:eval}, including downloading huge
collections of images tagged ``monitor'' from Flickr. In the end, we
followed Oquab~\cite{Oquab14} et al. and started with a model
pre-trained on the huge ImageNet dataset, even though that dataset has
nothing to do with lifelogging or monitor detection. Using those
network parameters as initialization, we then trained a network on monitor detection
using our relatively small training dataset. The exact mechanism that
allows this technique to work is not well understood, but may be that
there is enough common visual structure in the world that a neural network
trained for one problem still learns useful low-level features that also
apply to other seemingly unrelated problems.

For our implementation, we use the open-source Caffe deep learning
software~\cite{caffe}.  Minimal preprocessing is necessary in order to
use Caffe. Each image is downsampled such that the short axis is 256
pixels long. The center of the image is sampled along the long axis to
offer a 256x256 pixel image to the network. 

\subsection{Classifying applications on computer screens}
While detecting the presence of a computer screen alone may be useful
in some applications, access control policies that apply restrictions
to \emph{all} images with a monitor may be overly aggressive.  Thus,
we seek a method that discriminates amongst screen content at the
granularity of the application that is being used.  While what
constitutes `sensitive' image content is subjective and likely differs
from user to user, there are certain categories of applications that
display information that most people would find sensitive.  In this
paper we consider three categories: email applications, social media
websites, and instant messenger services. This is by no means an
exhaustive list, but provides a starting point for evaluation.

The system must handle images of screens that contain sensitive
applications but not necessarily when the quality of 
the image does not effectively resolve enough sensitive information.
For instance,  an image of a monitor
displaying a very sensitive email is not actually sensitive if the
camera is so far away that text cannot be resolved.
Thus, during our evaluation in Section \ref{sec:eval} we
address how well the classifier performs with respect to screens that
contain intelligible information. While further work is needed in determining what types of information are `unresolvable' under which conditions in general (e.g., photos and video), we concentrate on the more specific problem of intelligible text.
 
As we did with the screen detection in the last subsection, we rely on
deep learning methods using CNNs. Application detection is a strictly
more difficult problem than monitor detection, because the system must
choose the correct of several possible applications in addition to deciding if there
is a monitor in the image at all. Also, appearance of some websites is highly
variable; for instance, Gmail offers customized background ``themes'' that
can dramatically alter its appearance, while different users' Facebook feeds
appear differently due to differences in ads, friend activity, languages,
browser settings, etc. We test the ability of the classifier to generalize
across these differences, even if the training algorithm has never seen any
images from a particular user's lifelog, in Section \ref{sec:eval}.

\subsection{ScreenTag: conveying the sensitivity of screens} 

In Section \ref{sec:intro} we described several methods for assigning
labels to images during or after photo
collection~\cite{markit,wdac,klemperer2012}. Here we propose to do
both: in addition to the post hoc processing of raw lifelogging images
that we discussed in the last two sections, we also consider marking
screens themselves with labels that could help ease the burden of
screen content classification. 

For example, a regular lifelogger could then install the ScreenTag
application on their home and office computers. ScreenTag displays a
machine-readable barcode in a corner of the screen, encoding
information about which applications are currently running on the
system. When processing lifelogging images, \sys uses the monitor
detection and application classification techniques presented above
but also scans for this special barcode. If the barcode is missing,
because the user captures another person's monitor, \sys may still
take the correct action as long as the system classifies the image correctly.
If the barcode is present, the visual recognition task is eased significantly,
and we hypothesize that there is a greater chance that \sys will correctly
handle the image. In an era of pervasive cameras, people may be sufficiently motivated to include such privacy signals on their screens.

While it would be possible to use out-of-band channels to communicate
this information while leaving the screen content unaltered, these
channels lack precision, e.g., by blocking images even when the screen is not
within the field of view of the camera. Consider a policy to prevent
the photography of emails. The system could use Bluetooth to inform
nearby lifelogging cameras that the email application is running, but
photos of an area nowhere near the computer would be assigned an
incorrect label.  Thus, we pursue an in-band method of a visual marker
that is rendered on the screen, which we call ScreenTag. This approach is unique when compared to the MarkIt and WDAC systems in that the annotation changes dynamically with screen content and the images are algorithmically labeled. 

\begin{figure}[t]
\begin{center}
\fbox{\includegraphics[width=3.35in]{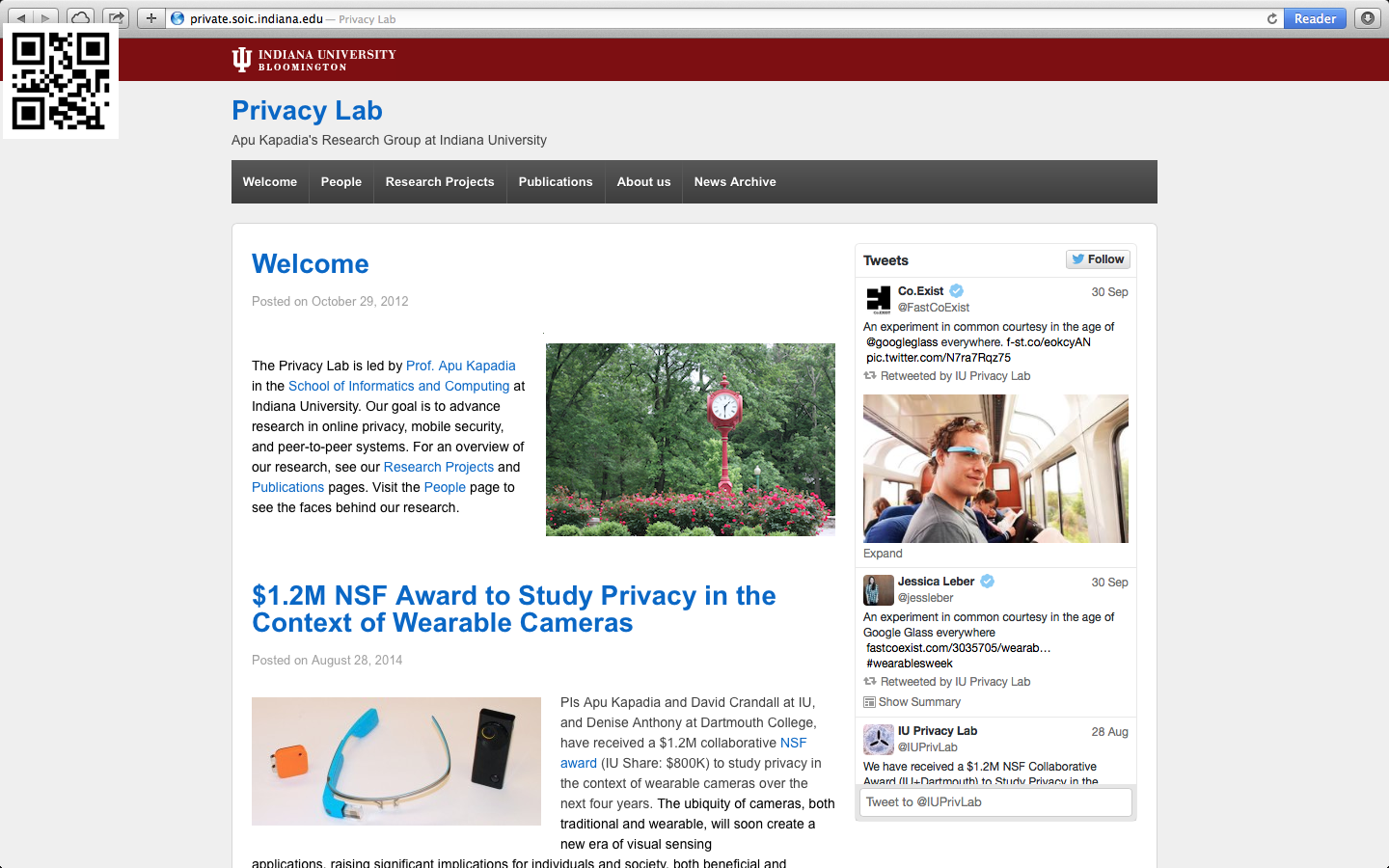}}
\end{center}
\caption{A screen capture with the ScreenTag visible in the upper left corner. This display is 1440x900 pixels and the QR code is set to 120x120 pixels. For this screen configuration, ScreenTag requires just 1.11\% of the viewing area.}
\label{fig:screentagpic}
\end{figure}

We prototyped the ScreenTag system for Mac OS X 10.9. We constructed a
blacklist of sensitive applications and websites, 
including Gmail, Facebook, and Apple Messenger for our
evaluation. Our program polls system processes and the Safari web
browser every second via bash and AppleScript, and constructs 
a bit vector encoding
the state of these
applications and if blacklisted websites are on the front tab of the
browser. This vector is encoded in a QR code that is configured for
maximum readability and the highest level of error correction using
QRencode~\cite{qrencode}. We use the Geektools software package 
to display the gadget persistently while
providing the user the ability to resize or move the gadget at their
will~\cite{geektools}. Figure~\ref{fig:screentagpic} shows a
screenshot of ScreenTag running while a browser window is open.

%% file: eval.tex

\section{Evaluation}
\label{sec:eval}
We evaluated \sys through numerous experiments using a variety of image data to assess classifier accuracy and performance. We first describe the datasets that we used.

\begin{table*}[t]
  \centering  
  \caption{An overview of the datasets that were used to evaluate machine learning approaches. The \emph{irb study} dataset is an aggregation of images from 36 users. The dataset from \emph{author} was collected by the authors from their own lifelogging devices. The \emph{flickr} images were manually scraped from Flickr and randomly sampled.}
{
\begin{tabular}{cccccc|c}

 & Facebook & Gmail & Messenger & other & no monitor & total \\
\midrule
irb study data & 35 & 12 & 2 & 736 & 1957 & 2742 \\
author& 2750 & 2659 & 3046 & 3749 & 6594 & 18798\\
flickr & 0 & 0 & 0 & 784 & 0 & 784 \\
\midrule
total & 2785 & 2671 & 2799 & 5269 & 8551 & 22319 \\
\end{tabular}
}
  \label{tab:data}
\end{table*}

\subsection{Evaluation datasets}
\label{subsec:data}
In our search for suitable evaluation datasets, we came across none
that were within the public domain. We sampled the lifelogs of the
authors as our primary source of data for our machine learning
approaches. The lifelogging devices used were a combination of Google
Glass, Narrative Clip, Autographer, and lanyard worn smartphones with
continuous photography applications. In all, the authors provided more
than 18,000 images that were manually labeled. The authors' IRB office
was consulted and this effort was deemed to not be human subjects
research.

To augment our data, Roberto Hoyle at Indiana University made a subset of their 2014 UBICOMP~\cite{hoyle2014privacy} dataset available to us and we secured the necessary IRB permissions. This dataset is very valuable in that it was collected in situ by 36 participants in a human subject study.

Lastly, we scraped more than 784 manually labeled images from Flickr to bolster our dataset. These images are screenshots that contain monitor content that are largely devoid of the physical monitor structure (e.g., bezels, logos, buttons, etc). Details of our datasets can be found in Table~\ref{tab:data}.

The \emph{irb} and \emph{author} datasets are actual lifelog image sets that were opportunistically collected under uncontrolled conditions. As such, photographic quality is generally poor with a significant fraction displaying poor composition, exposure, or focus. All sources of data were given an opportunity to delete very sensitive images that should not be part of the study. 

\subsection{Detecting computer screens and monitors}
\label{subsec:binary}

Our initial task is to evaluate the efficacy of a classifier to retrieve images with computer screens in them. To do this, we conducted three experiments:
\begin{itemize}
\item{\xhdr{Experiment Screen1}~- Train on 9,986 images from the \emph{author} training partition. Test the model on 1,842 \emph{author} images from the test partition that are randomly sampled such that there is an equal class distribution, so that a random classifier will achieve a baseline classification accuracy of 50\%.}
\item{\xhdr{Experiment Screen2}~- Train on 9,986 images from the \emph{author} training partition.  Test the model on all 2,742 \emph{irb study data} images. 28.6\% of these images have screens in them, which is the observed behavior from aggregating images from 36 users (so that a majority-class classifier will achieve a baseline accuracy of 71.4\%).}

\item{\xhdr{Experiment Screen3}~- Train on 9,986 images from the
  \emph{author} training partition.  Test the model on a mix of the
  1,958 \emph{irb} images without screens and 784 \emph{flickr} images
  with screens. This experimental test set replaces the \emph{irb}
  screen images with those scraped from Flickr (baseline remains
  71.4\%).}
\end{itemize}

\begin{table}[t]
  \centering  
  \caption{BVLC Reference CaffeNet pre-trained model configuration with modification for \sys. There are five sparsely connected convolutional layers and three fully connected layers that serve as a traditional neural network. Observe that only the last layer, fc3, changes with respect to the number of classes that are used. The parameter \emph{n} is equal to the number of classes.}
{
\begin{tabular}{l|cccc}
layer & \# of filters & depth & width & height \\
\midrule
data &  & 3 & 227 & 227 \\
conv1 & 96 & 3 & 11 & 11 \\
conv2 & 256 & 48 & 5 & 5 \\
conv3 & 384 & 256 & 3 & 3 \\
conv4 & 384 & 192 & 3 & 3 \\
conv5 & 256 & 192 & 3 & 3 \\
fc1 & 1 & 4096 & 1 & 1 \\
fc2 & 1 & 4096 & 1 & 1 \\
fc3 & 1 & n & 1 & 1 \\
\end{tabular}
}
  \label{tab:binary-network-config}
\end{table}

As described in Section~\ref{sec:overview}, we trained the
Convolutional Neural Network by starting with a model pre-trained on
the large ImageNet collection of Internet images. These network weights
are then used as initialization for a second round of training on 
our 9,986 \emph{author} life-logging training images. We use
the BVLC Reference CaffeNet pre-trained
model that is supplied with Caffe~\cite{caffe}. 
The network configurations for screen
and application classification are shown in Table
\ref{tab:binary-network-config}. The model has 2.3M neurons with
over 60M parameters. This reflects the memory limits of the
NVIDIA Tesla K20 processor that we used in our implementation (described in Section~\ref{subsec:performance}).

\begin{figure}[t]
\begin{center}
{\includegraphics[width=3.35in]{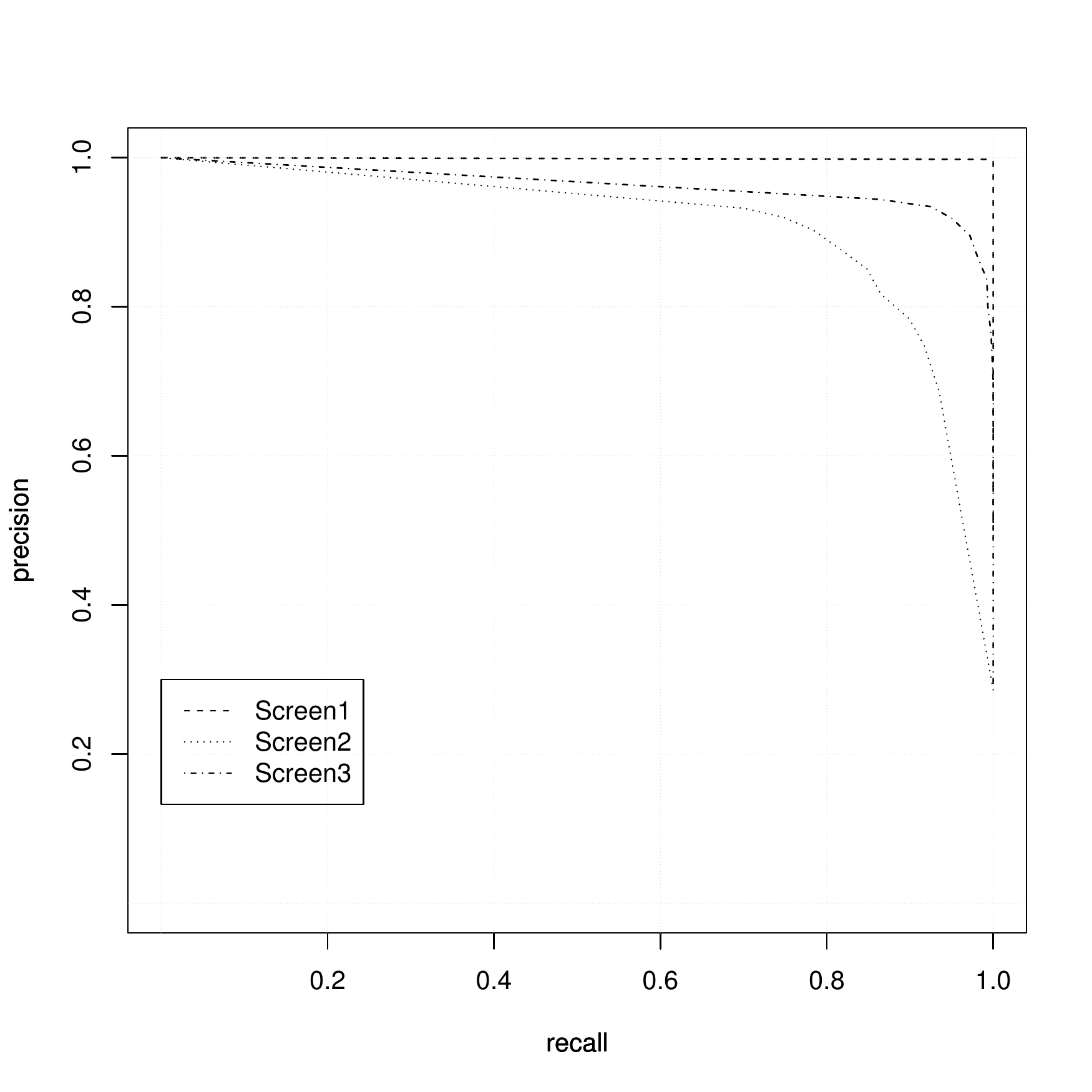}}
\end{center}
\caption{Precision and recall curves for retrieving images with computer screens.}
\label{fig:binary-prcurve}
\end{figure}

\xhdr{Experiment Screen1 results}~- This experiment is conducted to
serve as a sort of upper-bound on the accuracy for retrieving images
that have computer screens in them, because it is designed to be the
easiest of the experiments we consider. The algorithm must classify
unseen test image based on an independent set of training images, but
the training and test images are sampled from the same photo streams,
which means that there are likely to be very similar images in the two
sets.  The test partition was randomly subsampled to obtain an equal
class distribution~--- that is, a given image is just as likely to
contain a computer screen as it does not.

\begin{figure*}[t]
\begin{center}
\fbox{\includegraphics[height=1.1in]{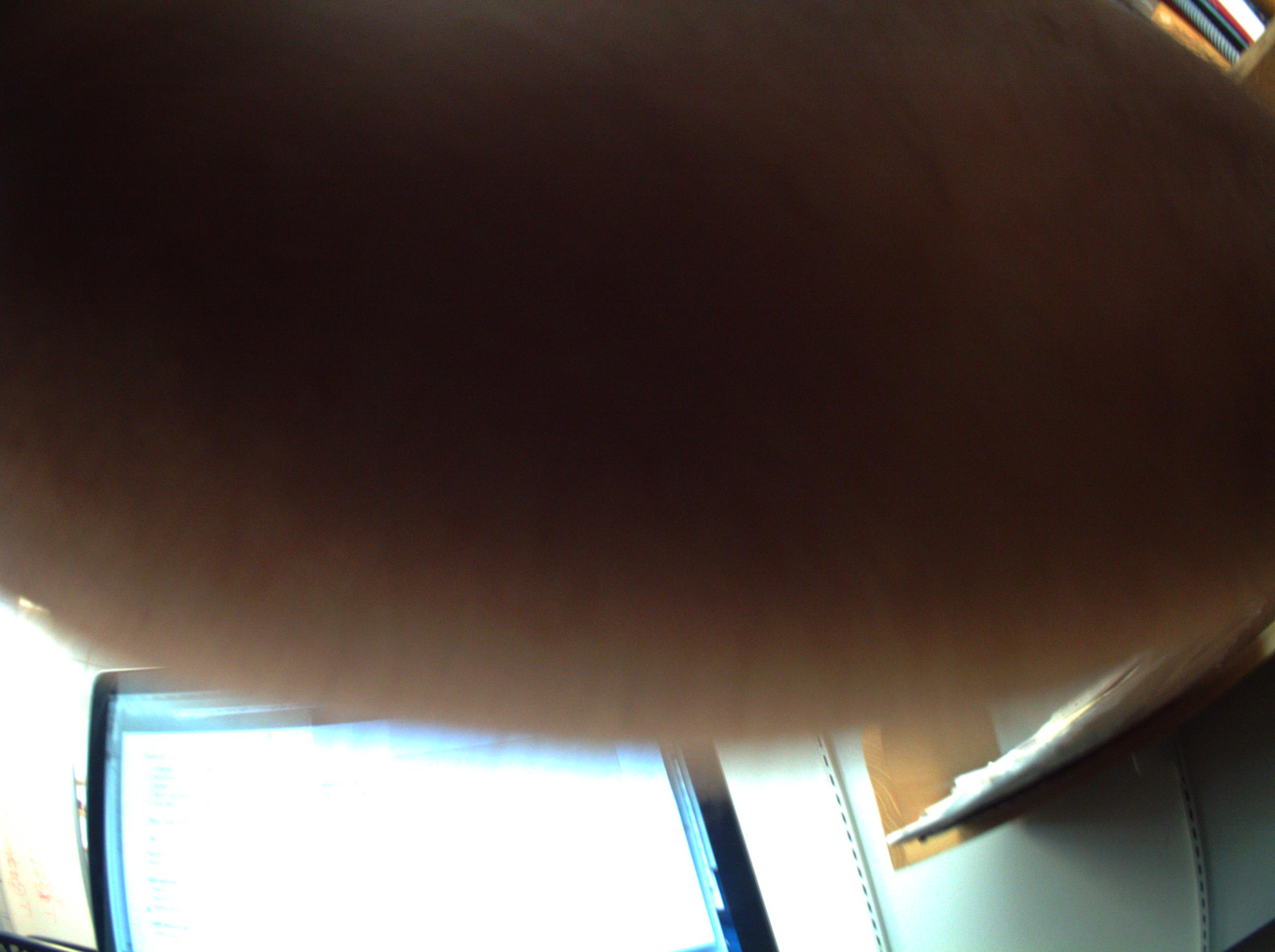}}
\\
\vspace{2mm}
\fbox{\includegraphics[height=1.1in]{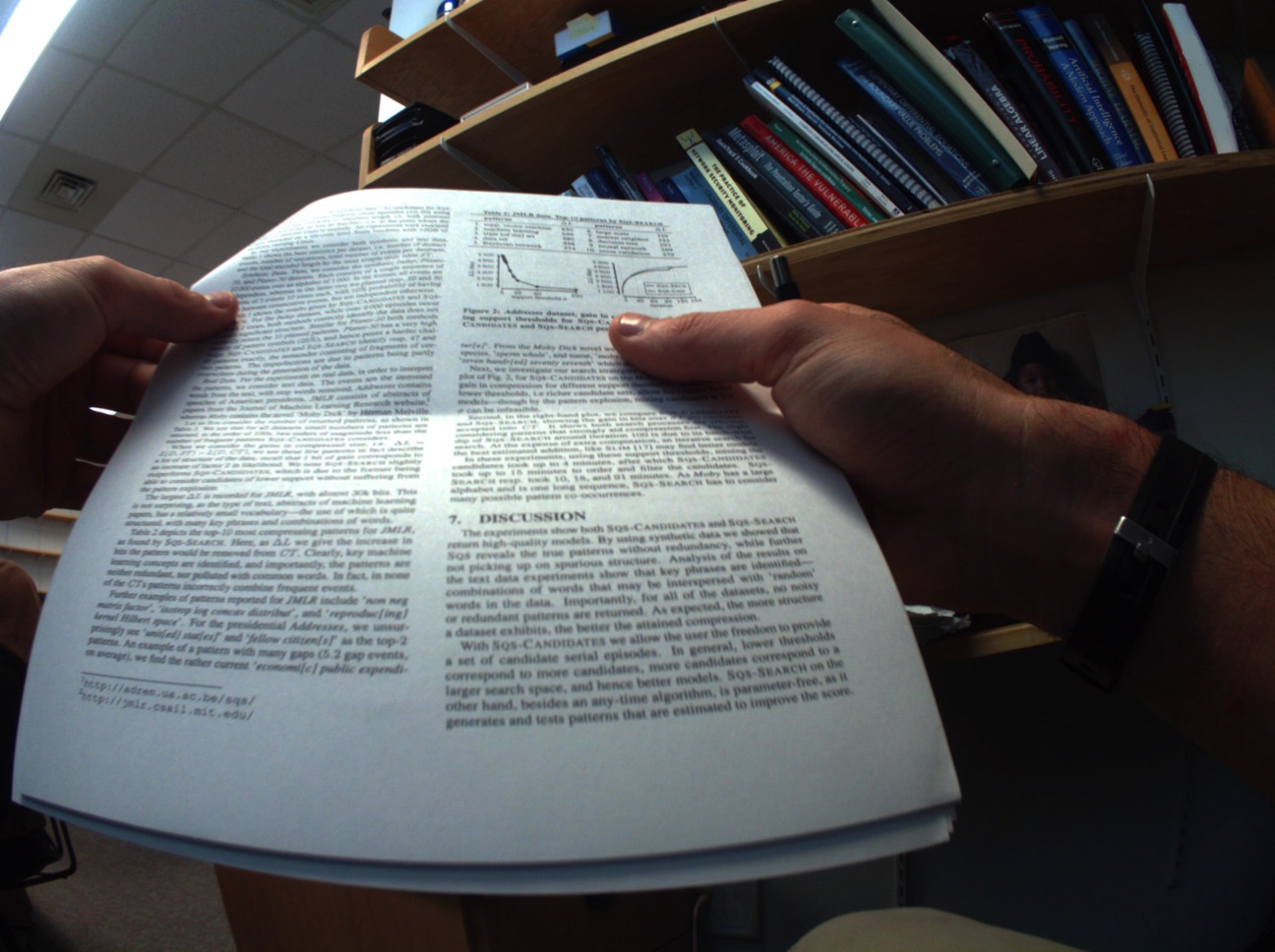}
\includegraphics[height=1.1in]{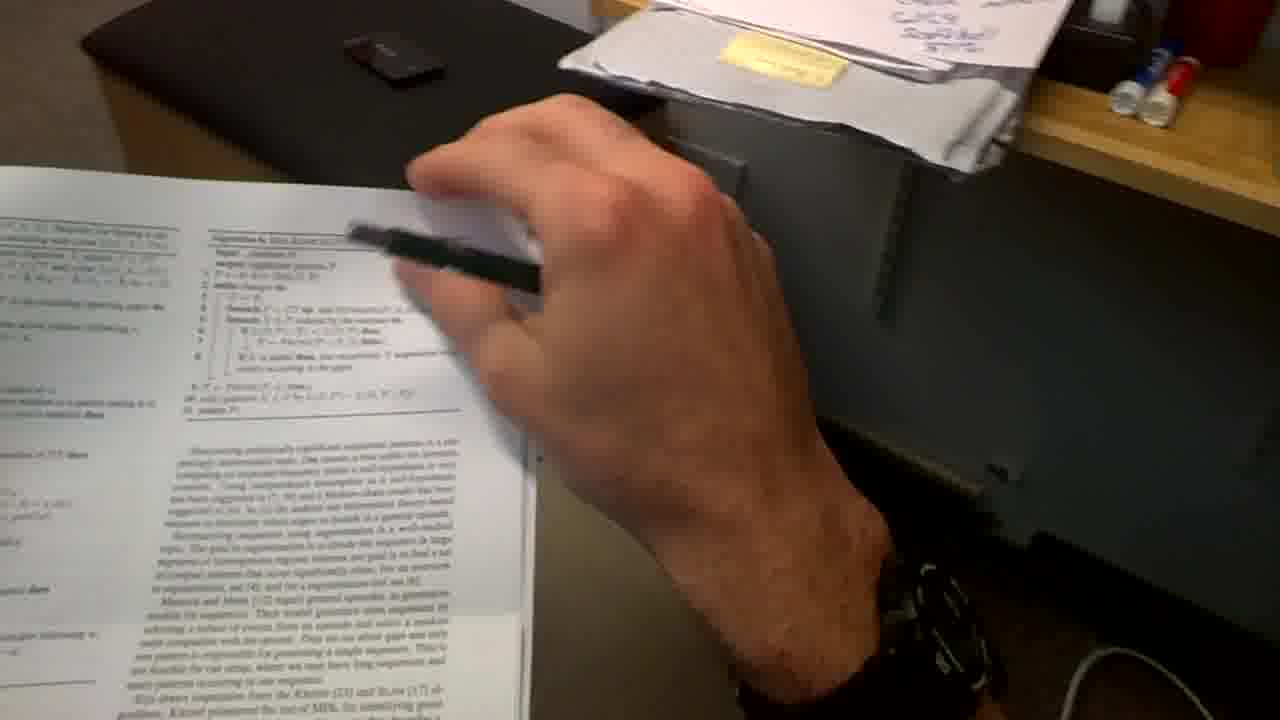}
\includegraphics[height=1.1in]{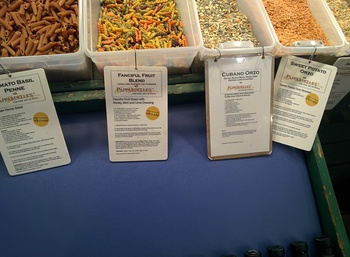}}
\end{center}
\caption{All four of the incorrectly classified Experiment \emph{Screen1} photos (there were 1842 images in this test set). The top panel contains the only false negative case which is mostly occluded with the screen over-exposed. The bottom panel contains the three false positive cases.}
\label{fig:misclass-examples}
\end{figure*}

The network demonstrated 99.8\% accuracy for this experiment. Table
\ref{tab:screen1-results} contains the confusion matrix that shows
only three false positives and one false negative. The incorrectly
classified images are displayed in Figure
\ref{fig:misclass-examples}. Observe that the sole false negative
image is of such poor quality that the no information can be retrieved
from the photographed screen (i.e., there would arguably be no
consequence if this image were classified incorrectly and
shared). The three depicted images that do not contain monitors are
labeled incorrectly and unnecessary restrictions would be applied in
our proposed use case.

Figure \ref{fig:binary-prcurve} shows this experiment cast as a retrieval problem for recalling images with screens in them. Performance is excellent with the ability to recall 99\% of screen images with 100\% precision.

\begin{table}[t]
  \centering  
  \caption{Experiment \emph{Screen1} confusion matrix. Baseline is 0.501. Accuracy is 0.998.}
{
\begin{tabular}{cccc}
&&predicted&\\
&&no screen&screen\\
\midrule
actual&no screen&919&3\\
&screen&1&919\\
\midrule
\end{tabular}
}
  \label{tab:screen1-results}
\end{table}

\xhdr{Experiment Screen2 results}~- This experiment tests the screen
classifier under more difficult conditions.  The test and training
datasets in this experiment are completely independent, because the
training images are from the \textit{author} dataset while the test
dataset is from the Hoyle et al. study, collected by 36 individuals in
unconstrained settings.  The class distribution in this case is not
balanced but instead reflects the true distribution of monitors
encountered in the real-world study, resulting in a high majority-class baseline.
Finally, the camera used to collect the test data is a Samsung Y smartphone
with software that is optimized to work under constrained battery
power and network bandwidth resources~\cite{hoyle2014privacy}. This
camera is not up to modern standards and as such, the images display
much higher degrees of motion blur, noise, and poor exposure (highlights).

The network demonstrated 91.5\% accuracy for this experiment. Table
\ref{tab:screen2-results} contains the confusion matrix that shows a
near equal mix of false negative and false positive instances. These
test images are IRB-controlled human subject study data so we are
unable to include them in this paper. However, we did manually review
all incorrectly classified images and report our observations.  Table
\ref{tab:screen1-fn-analysis} provides an analysis of the 117 false
negative images. In Section \ref{sec:intro} we speak to the challenge
of classifying computer screens that render content that looks unlike
computer applications. This table shows that 49.6\% of the false
negative images had computer screens present that were displaying
video games in full screen mode. Interestingly, the game Minecraft
represented a large fraction of these. About 12.8\% of the images capture media in full screen mode (movies, sports, and
television shows). It is important to note that the training data had
no examples of these types of images.

To assess the privacy impact stemming from classifier performance, we
seek to identify false negative images that do in fact have sensitive
content that would be potentially leaked. We found a total of 8 images
that contained sensitive content by a conservative definition (1
Skype screenshot, 2 Microsoft Word screenshots, 3 Facebook shots, and 2 Adobe Illustrator shots). This
represents a small fraction of the false negatives (6.8\%) and only
0.3\% of the overall test images.

We also manually reviewed the false positive images, and the results
are presented in Table \ref{tab:screen1-fp-analysis}. A significant
source of false positive instances came from images where windows or
other framed objects were prominent. A key feature of computer screens
is the boundary or frame that borders the display~--- this shows the
reliance of the classifier on invariant screen frames versus the
contents within. Additionally, about 16.4\% of the false positive
images were screens of televisions, projectors, or smartphones instead
of computers.  This is not necessarily an ill-effect because these
displays also often display private information, and demonstrates the
semantic power and the generalizability of deep learning techniques.

The results are plotted in a PR curve in Figure
\ref{fig:binary-prcurve}. As expected, the results are significantly
worse than the \emph{screen1} experiment, but even in this difficult
test case we are able to retrieve 88\% of screen images with 80\%
precision and observe adequate performance.

\begin{table}[t]
  \centering  
  \caption{Experiment \emph{Screen2} confusion matrix. Baseline is 0.714. Accuracy is 0.915.}
{
\begin{tabular}{cccc}
&&predicted&\\
&&no screen&screen\\
\midrule
actual&no screen&1842&117\\
&screen&116&667\\
\midrule
\end{tabular}
}
  \label{tab:screen2-results}
\end{table}

\begin{table}[t]
  \centering  
  \caption{Experiment \emph{Screen2} false negative (FN) analysis. The FN images were manually reviewed and the following observations were made about the listed fraction of images. We speculate that these observed properties frustrated classification attempts. Note that these observation categories are not mutually exclusive.}
{
\begin{tabular}{lr}
&fraction of FN images\\
\midrule
full screen video games&0.496\\
less than 50\% of screen visible&0.489\\
significantly out of focus&0.350\\
movie or TV show being played&0.128\\\\
screen with sensitive information&0.003\\
\midrule
\end{tabular}
}
  \label{tab:screen1-fn-analysis}
\end{table}

\begin{table}[t]
  \centering  
  \caption{Experiment \emph{Screen2} false positive (FP) analysis. The FP images were manually reviewed and the following observations were made about the listed fraction of images. We speculate that these observed properties frustrated classification attempts. Note that these observation categories are not mutually exclusive.}
{
\begin{tabular}{lr}
&fraction of FP images\\
\midrule
prominent window visible&0.336\\
other \emph{framed} element&0.328\\
non-computer device with screen&0.164\\
\midrule
\end{tabular}
}
  \label{tab:screen1-fp-analysis}
\end{table}

\xhdr{Experiment Screen3 results}~- 
In this experiment, we test the ability of a classifier trained
on one type of images to classify images of another type.
This experiment is related to
experiment \emph{Screen2} in that they share the same negative class
images (those without monitors), but the positive class  contains monitor
images that are randomly collected from Flickr and largely consists of
screenshots of applications, not lifelogging photographs of
screens. The difference is that here we are presenting the classifier
with screen content sans computer monitor features (e.g., bezels,
computer screen logos, etc).

The classifier had an improved accuracy of 95.3\%, which was achieved by reducing the false negative rate when compared to experiment \emph{Screen2}. The confusion matrix can be found in Table \ref{tab:screen3-results}. For this experiment, the PR curve in Figure \ref{fig:binary-prcurve} shows that we recall 98\% of screen images with 80\% precision. This experiment further demonstrates the ability to detect monitors in general.

\begin{table}[t]
  \centering  
  \caption{Experiment \emph{Screen3} confusion matrix. Baseline is 0.714. Accuracy is 0.953.}
{
\begin{tabular}{cccc}
&&predicted&\\
&&no screen&screen\\
\midrule
actual&no screen&1842&117\\
&screen&12&771\\
\midrule
\end{tabular}
}
  \label{tab:screen3-results}
\end{table}

\subsection{Classifying applications}
\label{subsec:multiway}

While coarse policies that act solely on the presence of screens in images offer utility, these may be overly restrictive. That is, there may be nonsensitive images that users desire to share. Thus, we seek to classify images further based on screen content. We do this on the basis of applications that render content on the display. To evaluate \sys in this manner, we conducted the following three experiments:

\begin{itemize}
\item{\xhdr{Experiment App1}~- Binary classification between
  \emph{sensitive} applications versus other applications. Train on 9,986 images from the \emph{author} training partition. Test the model on 5,050
  \emph{author} images from the test partition that are randomly
  sampled such that there is an equal class distribution (baseline is
  50\%).}

\item{\xhdr{Experiment App2}~- Four-way classification between
  Facebook, Gmail, Apple Messenger, and an ``other'' category. Train on 9,986 images from the \emph{author} training partition. Test the model on 6,868
  \emph{author} test images sampled for an equal class distribution
  (baseline is 25\%).}

\item{\xhdr{Experiment App3}~- Five-way classification between
  no-screen, Facebook, Gmail, Apple Messenger, and an ``other''
  application category. Train on 9,986 images from the \emph{author} training partition. Test the model on all 2,742 \emph{irb study data}
  images. 28.6\% of these images have screens in them, which is the
  observed behavior from aggregating images from 36 users (baseline is
  71.4\%). The distribution of other applications is extremely
  unbalanced as shown in Table~\ref{tab:app3-results}.}
\end{itemize}

For these experiments with increased numbers of classes, we modify only the last layer of the convolutional neural network as shown in Table \ref{tab:binary-network-config}.

\begin{figure}[t]
\begin{center}
{\includegraphics[width=3.35in]{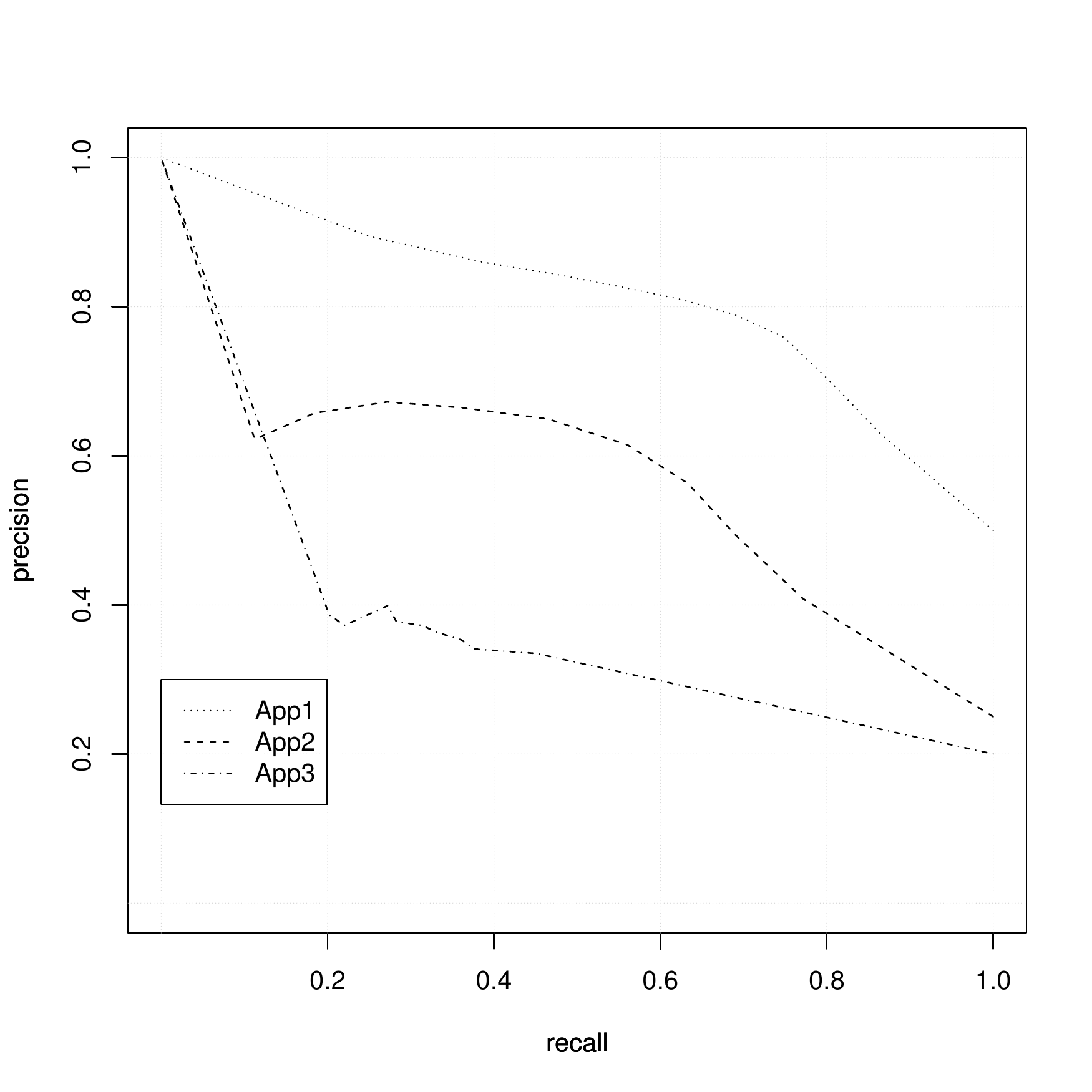}}
\end{center}
\caption{Precision and recall curves for the application classification experiments.}
\label{fig:app-prcurve}
\end{figure}

\xhdr{Experiment App1 results}~- This experiment expresses application classification as a binary task~--- a `sensitive application' class includes images from Facebook, Gmail and Apple Messenger while an `other application' class applies to screens displaying anything else. 

The classifier demonstrates an accuracy of 75.1\% which is 50\% better than randomly guessing whether an image is sensitive or not. Table \ref{tab:app1-results} shows the confusion matrix which interestingly shows that the classifier has a greater bias for false positives than false negatives. That is, the classifier is more likely to be overly restrictive by labeling `other applications' as sensitive than vice versa. The PR curve in Figure \ref{fig:app-prcurve} shows that this classifier can recall 80\% of sensitive applications with 71\% precision.

\begin{table}[t]
  \centering  
  \caption{Experiment \emph{App1} confusion matrix. Baseline is 0.500. Accuracy is 0.751.}
{
\begin{tabular}{cccc}
&&predicted&\\
&&other app&sensitive app\\
\midrule
actual&other app&1717&808\\
&sensitive app&449&2076\\
\midrule
\end{tabular}
}
  \label{tab:app1-results}
\end{table}

\xhdr{Experiment App2 results}~- We now seek to determine the performance of a classifier that attempts to discriminate amongst individual applications. Such fine-grained discrimination enables more expressive policies that could for example allow a user to wholly restrict images taken of their email application while allowing them to share images of their social media applications with friends.

The network was able to classify the test images with an accuracy of 54.2\%.  While this is degraded from the previous binary classification case, the baseline is similarly decreased to 0.25. Table \ref{tab:app2-results} contains the confusion matrix for this experiment. This shows that the classifier is much more likely to label `other applications' as Apple Messenger than it is to label Messenger images as an 'other application' on our dataset. But, this also shows that the classifier undesirably labels both Facebook and Gmail images as `other applications' more often than vice versa. 

\begin{figure*}[t]
\begin{center}
\includegraphics[width=2.2in]{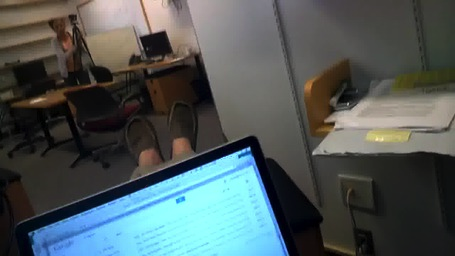}
\hspace{3mm}
\includegraphics[width=2.2in]{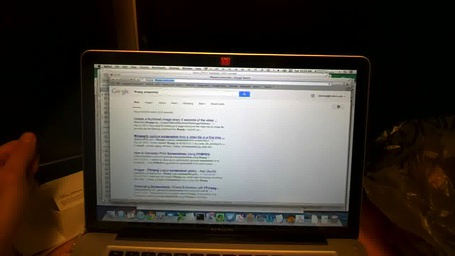}\\
\hspace{.6in}Gmail \hspace{1.2in} \emph{other} app (Google search)\\
\vspace{4mm}
\includegraphics[width=2.2in]{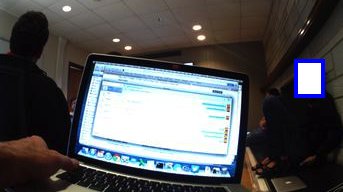}
\hspace{3mm}
\includegraphics[width=2.2in]{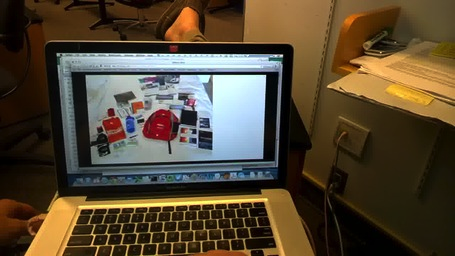}\\
Messenger \hspace{1.5in} Facebook\\
\end{center}
\caption{Examples of images that were correctly classified in experiment \emph{App2}. Note the ability of the classifier to discriminate amongst Google search and GMail which have similar visual features. The blue box is added for anonymity.}
\label{fig:appclass-examples}
\end{figure*}

The same table also shows the inter-app confusion. While the performance is not good, we can look to some example images to see how the classifier performs. Figure \ref{fig:appclass-examples} contains an example image from each of the four categories that was classified correctly. Observe that the classifier was able to distinguish between Google search and Gmail even when they contain similar visual features. The correctly classified Facebook image that is shown displays a picture in a mode where the expected blue Facebook banner is absent~--- it would be a challenge for the typical user to accurately label the application in this case.
\newpage

We carefully chose the representative applications that we did in order to rigorously evaluate \sys:
\begin{itemize}
\item{Facebook displays a large degree of variation in visual content. Signature visual features (e.g., the blue banner) come and go depending on context. Much of the screen contains content personalized to the user.}
\item{Gmail is an example of an email service that is browser-based and difficult to visually distinguish from other web content (especially other Google web services).}
\item{Apple Messenger has a minimalist visual theme that was deliberately chosen as an example of a messenging application that is not easily recognizable.}
\end{itemize}

It is intuitive that \sys's ability to discriminate amongst a given pair of applications is largely dependent on the choice of applications. Our evaluated applications and lifelogging datasets present challenging cases and would expect improved performance in the general case.

The \emph{screen2} PR curve shown in Figure \ref{fig:app-prcurve} demonstrates a degradation of retrieval performance as compared to the \emph{screen1} curve, since we seek to make an already difficult problem even more challenging. The classifier in this case can recall 80\% of the desired images with a precision of less than 40\%. 

\begin{table*}[t]
  \centering  
  \caption{Experiment \emph{App2} confusion matrix. Baseline is 0.250. Accuracy is 0.542.}
{
\begin{tabular}{cccccc}
&&predicted&\\
&&other app&messenger&facebook&gmail\\
\midrule
actual&other app&1165&378&24&150\\
&messenger&148&1403&0&166\\
&facebook&379&635&524&179\\
&gmail&460&602&23&632\\
\midrule
\end{tabular}
}
  \label{tab:app2-results}
\end{table*}

\xhdr{Experiment App3 results}~- Lastly, we consider an experiment
that reflects more difficult conditions, by introducing data with
five classes, including four application
classes and the case that there is no
screen in the image. While our \emph{author} training data has reasonably balanced
classes, the \emph{irb study} test data for this experiment has a high
degree of imbalance.

The resulting accuracy for this experiment is 77.7\% which is marginally above the 0.714 baseline. Thus, in this case the classifier cannot do much better than random guessing. The confusion matrix is displayed in Table \ref{tab:app3-results}. We can see that the classifier performs well at the coarse level of inferring whether or not a screen is present, but classification amongst sensitive applications is very poor. We conclude this subsection with the PR curve shown in Figure \ref{fig:app-prcurve}. This classifier is able to retrieve 80\% of desired images with a precision of about 25\%.

\begin{table*}[t]
  \centering  
  \caption{Experiment \emph{App3} confusion matrix. Baseline is 0.714. Accuracy is 0.777.}
{
\begin{tabular}{ccccccc}
&&predicted&\\
&&no screen&other app&messenger&facebook&gmail\\
\midrule
actual&no screen&1882&59&6&0&12\\
&other app&157&243&143&35&158\\
&messenger&0&2&0&0&0\\
&facebook&4&12&11&5&3\\
&gmail&0&7&3&0&2\\
\midrule
\end{tabular}
}
  \label{tab:app3-results}
\end{table*}

\xhdr{Other application classification approaches}~- Given the demonstrated difficulty of application classification, we explored other experiments outside of the three that we detail above.

An advantage of using CNNs is in the manner by which they extract useful features in the convolutional layers~--- thus, we consider using CNN-generated features with a different choice of classifier.  We extracted the features from the network and applied them to SVM classifiers~\cite{REF08a}.  We applied two models to extract the features: the standard BVLC Reference CaffeNet pre-trained model and the fined-tuned model based on our data set (the latter case coincidentally represents the features used internally to the CNN in \emph{App1}, \emph{App2}, and \emph{App3}).
 However, these attempts end up being inferior to the neural network classifier that is provided by Caffe.

\subsection{ScreenTag performance}
\label{subsec:screentag}

We evaluated our ScreenTag system by running it as described in Section \ref{sec:overview}. We defined a set of monitored applications and websites (Facebook, Gmail, and Apple Messenger) and ran our ScreenTag service to persistently display the QR code marker in the upper-left corner of the 1440x900 screen at a size of 120x120 pixels as shown in Figure \ref{fig:screentagpic}. In this configuration on the test machine, ScreenTag covers 1.11\% of the viewable screen area. The system is configured to update the marker at a rate of 1Hz. We invoke the highest level of error correction, H, to improve the readability of the QR code~\cite{qrstandard}. In theory, this allows the code to be read with nearly 30\% of the visual information missing.

\begin{table}[t]
  \centering  
  \caption{ScreenTag results.}
{
\begin{tabular}{ccc}

fraction of&\# of images& \% of \\
ScreenTag visible&(\%)&ScreenTags read\\
\midrule
full&511&89.6\\
partial&11&0.0\\
none&13&0.0\\
\midrule
TOTAL&535&85.6
\end{tabular}
}
  \label{tab:screentag-results}
\end{table}

\begin{figure*}[t]
\begin{center}
\fbox{\includegraphics[width=2.2in]{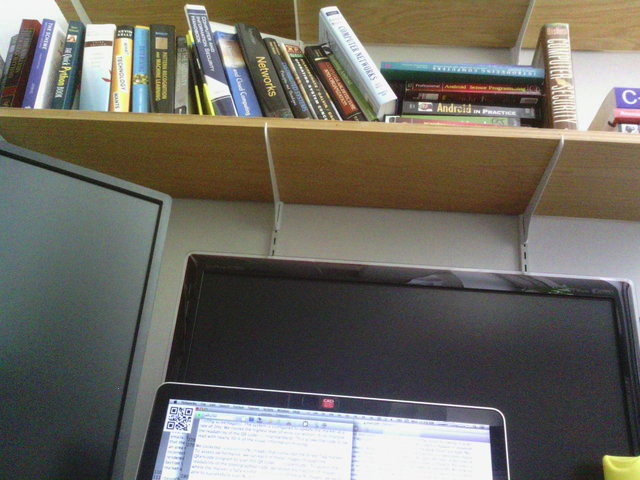}
\hspace{3mm}
\includegraphics[width=2.2in]{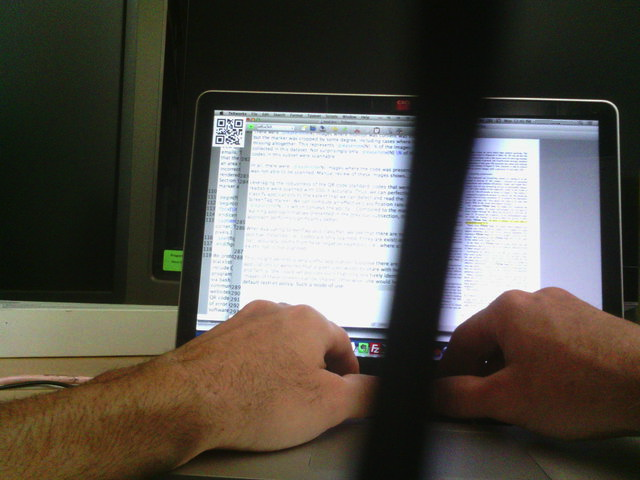}}\\
ScreenTag \emph{was} successfully scanned\\
\vspace{4mm}
\fbox{\includegraphics[width=2.2in]{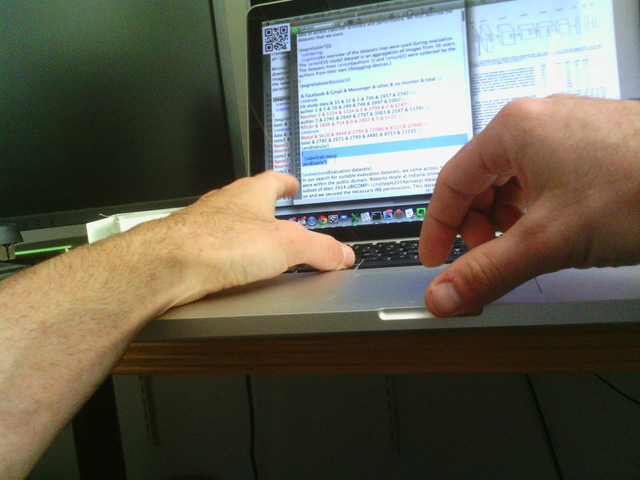}
\hspace{3mm}
\includegraphics[width=2.2in]{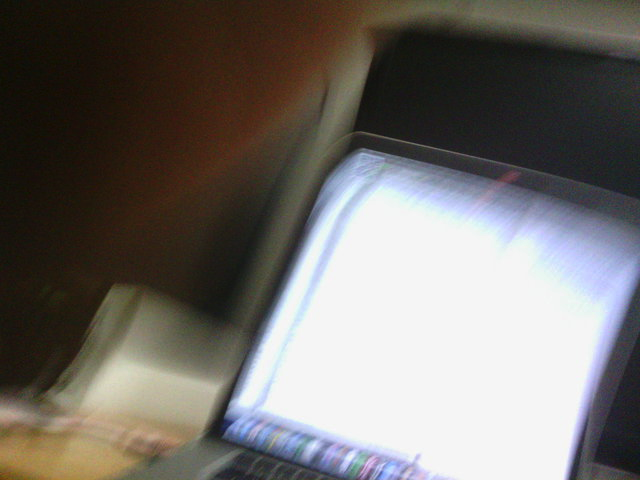}}\\
ScreenTag was not scanned\\
\end{center}
\caption{Examples of images where ScreenTag is rendered on displays. Observe that bottom left image has sufficient resolution and sharpness to reveal the text on the screen. The example on the bottom right has a large degree of motion blur so neither the QR code nor anything else can be interpreted.}
\label{fig:screentag-examples}
\end{figure*}
\newpage
We collected 535 images while using a laptop computer with ScreenTag rendered. To assess performance, we ran each of these images through the open source ZBar program to scan the QR code~\cite{zbar}. The results are shown in Table \ref{tab:screentag-results}. We first seek to understand the readability of photographed codes in cases where the QR code is fully visible (no cropping or occlusion). We find that of these 511 images, we were able to successfully scan 85.6\%.

There were 24 (4.5\%) images where monitor was content was visible, but the marker was cropped by some degree, including cases where it was missing altogether. None of the codes in this subset were scannable, even those codes that were cropped by less than the 30\% that the error correction should have recovered. While the error correction in QR codes adds a layer of robustness, our codes are scanned from images that are taken from some distance away with noise, illumination, and rotation transforms applied.

In all, there were 64 images where the ScreenTag was present, but was not able to be scanned. Manual review of these images shows 13\% of these had such a high degree of poor exposure and focus that nothing on the screen was intelligible. Figure \ref{fig:screentag-examples} shows examples of challenging images where ScreenTag was read correctly and examples of those images that could not be scanned.

Because of the built-in robustness of the QR code standard,
codes that were
readable were scanned with 100\% accuracy. Thus, we can perfectly
classify applications to the extent that we can detect and read the
ScreenTag marker. The effective classification rate of
89.9\% means it  performs significantly better than the five-way
application classification results of experiment \emph{App3} in
Subsection \ref{subsec:multiway}. When considering screen images, our
experimental baseline is 0.25 with only 2 bits of information
encoded in the QR code. A version 1 QR code allows the encoding of 72 data
bits, so ScreenTag has the ability to discriminate amongst a
\emph{much} larger number of applications while retaining the same
accuracy.

When evaluating ScreenTag as a classifier, we see that there are no false positive instances (i.e., codes are only scanned if they are exist) and that our error stems from false negative examples (i.e., where a code exists, but is not scanned).

This insight permits a very useful application. Suppose there are applications or websites that a given user wants to share with their friends and family. They could set policies such that only positively identified images of these screens can be shared. Otherwise, they would have a default restrict policy. Such a mode of use could act in a privacy preserving way so long as we trust the system to not render a ScreenTag that marks private information as something to be shared. Consider our running example:

\begin{quotation}
\emph{Mary decides that she only wants to share her screen images while playing Minecraft and while using her illustration application. She configures ScreenTag to mark her screen when she is using these applications and creates a \sys policy that allows these pictures to be shared.}
\end{quotation} 

We limited our evaluation to the single marker size and location, but other options are possible. Adding
additional markers and increasing its size should increase the
likelihood of successful scans at the expense of a further reduction
in usable screen space. Furthermore, even a version 1 QR code allows
more capacity than may be necessary for our application. A bespoke
code configuration could decrease data density in order to improve
readability. We reserve this additional evaluation for future work.

\subsection{Computational performance}
\label{subsec:performance}

For the machine learning approaches that we presented in Subsections \ref{subsec:binary} and \ref{subsec:multiway}, we used a workstation with an AMD Opteron 16-core Interlagos x86\_64 CPU processor and one NVIDIA Tesla K20 GPU accelerator with a single Kepler GK110 GPU. The Caffe implementation ran on the single GPU. We began with the BLVC Reference CaffeNet model so we only had to fine-tune the network with our labeled training images. For the experiments described in Subsections \ref{subsec:binary} and \ref{subsec:multiway}, the training period ranged from 3 to 5 hours. However, classification computation time for individual images was just 0.12 seconds on average to include preprocessing and oversampling steps. The same classification task on the CPU averaged 1.5 seconds per image, which validates that it is feasible that computation can be performed on the users' machines in order to avoid relying on an untrusted cloud.

The ScreenTag system involves a much less computationally-intensive task. On average, it took just 0.44 seconds for the ZBar program to scan the image using an Ivy Bridge i7 laptop. This means that it is feasible that images can be curated in real time by the collection device when screens are annotated.

\newpage

%% file: discussion.tex

\section{Discussion}
\label{sec:disc}

\xhdr{Thwarting the photography of screens.} As discussed in Section
\ref{sec:intro} we spend a large fraction of our time in front of
computer screens engaging in private communications, conducting
business among other sensitive functions. The confluence of our uses
of portable computing and wearable cameras creates an environment
where we can conduct these functions almost anywhere while within the
view of others.

While we focus on photography of screens, a related vulnerability exists if the person sitting nearby at the coffee shop can read a private email directly from your screen. Systems have been proposed that seek to identify people~\cite{Ali:2014:PMU:2638728.2638788} looking at your screen, but  a motivated attacker with inexpensive magnification devices could still leave a victim vulnerable in public settings. The problem is worse when attackers employ camera devices. One system seeks to identify and disable nearby cameras~\cite{Truong:2005:PCR:2156905.2156910}, but prior work has shown  powerful attacks on our screens where the attacker is up to 50 meters away and only views a reflection of your monitor~\cite{DBLP:conf/ccs/RaguramWGMF11,xu2013seeing}.

A different approach is to design the screen and content in such away
that undesired viewing and photography is made difficult. This can be
done by using a physical filter that is placed over the screen to
restrict the possible viewing angle~\cite{3m-privacy} or by creatively
engineering the screen content. For instance, the Yovo messaging application renders
screen content in a highly dynamic way, such to make static photography more
difficult~\cite{yovo}.

The lifelogging mode of use, made possible with modern wearable camera devices, begs for different solutions. The Hoyle study shows that continuous opportunistic photography represents a privacy threat to the users of the devices and those that are around them~\cite{hoyle2014privacy}, but without malicious intentions.
\sys is a system that allows lifeloggers to more easily curate their vast collections of images in a privacy preserving way. Absent approaches to keep pictures of screens out of our lifelogs, we provide a manner in which to handle them with usable policies.

\xhdr{Avoiding the photography of bystanders' screens.} Concern of other people's privacy out of a sense of propriety is a subset of our problem. The previous discussion focused on potential images of screens from the perspective of the bystander. Here, we consider lifeloggers collecting images in the presence of strangers using their electronic devices. The coarse screen detector component of \sys remarkably labels images with screens in them agnostic of content.

While outside of the scope of this work, it is also possible that bystanders can communicate policies about their screens in the WDAC schema~\cite{wdac}, which is similar to the work of Schiff et al. where bystanders wear visible markers to communicate policies to surveillance cameras~\cite{schiff2009respectful}.
However, our system does not easily differentiate between our own screens and devices that belong to bystanders. Thus, propriety policies for bystanders and default sharing policies of our own screens are contradictory if dependent on the same attribute. We reserve further exploration of these solutions for future work.

\xhdr{Screens signaling sensitive information.} The ScreenTag system leverages the QR code which benefits from a well-embraced standard with demonstrated success in many applications. Our application of it permits the transmission of a significant amount of data about screen context. An alternative approach might use a bespoke method of communicating visual information to cameras. This can be done using some sort of rendered watermark~\cite{hilight} or with visible elements that provide informative features to machine learning approaches.

\begin{figure}[t]
\begin{center}
{\includegraphics[width=1.3in]{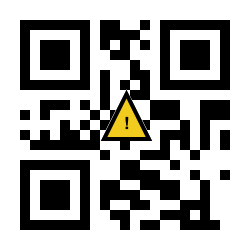}
\hspace{6mm}
\includegraphics[width=1.3in]{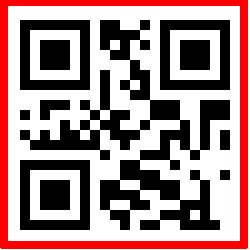}}
\end{center}
\caption{A standard QR code may be modified to add visual elements to convey information to the user. Consider these two examples that might signal a sensitive context. Error correction permits the modification of the code itself to a degree as the code on the left is still readable.}
\label{fig:othertags}
\end{figure}

Our ScreenTag system is limited in that it only provides machine readable information. While it is effective in communicating contextual information to cameras, users may benefit from knowing when their screen has sensitive information that requires judicious behavior. An added feature could display a marker that is visible to users that dynamically changes based on screen content. A motivating example is the classification banner that is rendered on computer displays on DoD information systems~\cite{dod-classbanner}. We envision that visual elements can be added to our existing QR code to let users know they are using an application that is especially sensitive. We offer examples in Figure \ref{fig:othertags} and save further evaluation for future work.

\xhdr{Usability.} Machine learning techniques require some degree of training data. For instance, the PlaceAvoider system required that the user `enroll' their spaces to create labeled training data~\cite{placeavoider14ndss}. This approach is not feasible for \sys~--- our deep learning-based system benefits from copious numbers of labeled images (on the order of thousands of images or more). 
Our results show that \sys offers good general performance using a limited training set of less than 10,000 images sourced from just two users. The performance stands to increase with a richer source of training data. 
A usable \sys application would leverage existing trained models that users could benefit from immediately only having to define policies.

While the screen detection algorithm performed extremely well, the application classification task suffered from a high degree of error to the extent that reliance on the classifier for policy enforcement is not prudent. More work remains to be done in this area. However, the ScreenTag system performed well at discriminating amongst different applications. The ScreenTag approach allows the user to balance performance and usability by defining the size and location of the marker.

As described in Section~\ref{subsec:performance}, the running time
of our classifier benefits significantly from having a GPU.
Advanced mobile devices like smartphones have GPUs that could be used for this purpose,
although current-generation wearable devices do not. 
However, \sys could
easily be implemented alongside the cloud-based services that
accompany our current lifelogging devices or though an OS-level cloud
service akin to Apple Siri~\cite{siri} or Google
Voice~\cite{googlevoice}. In addition to our privacy objectives, the
more general application of image tagging could serve to help users
curate their images.

%% file: related.tex
\section{Related Work}
\label{sec:related}

\xhdr{Lifelogging and privacy.}
The recent availability of wearable devices for consumers has resulted in even greater interest by the research community. Work by Hoyle et al. explores privacy issues for lifeloggers~\cite{hoyle2014privacy} while Denning et al. consider the issues of bystanders that find themselves in the vicinity of users of wearable cameras~\cite{Denning2014}. Roesner et al. address the general security and privacy issues for augmented reality devices which apply also to wearable camera devices~\cite{Roesner2013}. Caine explores mistakes that users make when they share information with an unintended group, a problem that \sys addresses~\cite{Caine2009}. 
The PlaceRaider system is a smartphone based attack that shows how opportunistically-collected images can be exploited by an adversary to reconstruct 3D models of their personal spaces~\cite{placeraider-ndss13}.  These works motivate the necessity of controls that can help users best collect and manage lifelogs. 

\xhdr{Access control.}
Discretionary- and mandatory-access control frameworks underpin many traditional computer operating systems~\cite{DepartmentofDefense1985}, but research on access control concepts for sensing platforms is bringing about new ideas. These sensor-enabled products include wearable and mobile devices that differ in how files (objects) are created and used. User-driven access control~\cite{Roesner:2012:UAC:2310656.2310678} seeks to add abstraction layers that confirm user intent before the system grants access to user-owned resources like cameras. World-driven access control (WDAC) reads policies that are embedded in the environment (e.g, tags on physical objects)~\cite{wdac}. Attribute-based access control (ABAC) collects or infers attributes from the environment from which to apply policy rules to~\cite{templeman-abac}.

Moreover, other systems perform access control functions for image objects by filtering content such that applications can only access necessary information. The Darkly system~\cite{scannerdarkly} and the related OS recognizers~\cite{recognizers} perform this function using computer vision approaches. \sys combines elements from each of these systems to offer a solution to the problem of screens that are photographed by wearable cameras. 

\xhdr{Object detection.}
Object detection is an  active field in the computer vision community. Great strides have been made in the last 10 years with the identification of useful visual features including SIFT~\cite{sift} and HOG~\cite{dalal05hog} that are used with traditional classifiers like support vector machines. Greater advancement was seen when newer approaches were applied to existing features~--- this includes the `bag of words' approach by Csurka et al.~\cite{csurka2004visual}. Chatfield et al. provide a comprehensive review of these techniques in their comparative paper~\cite{chatfield2011devil}

 Object detection competitions help to motivate new approaches and provide an objective measure by which to compare different algorithms. As described in Section \ref{sec:overview}, techniques based on deep learning with CNNs have  surpassed existing techniques. Again, Chatfield et al. provide a useful survey of this new technology~\cite{chatfield14}.

%% file: conclusion.tex

\section{Conclusion}
\label{sec:conclusion}

Research on the use of wearable cameras shows a need for solutions that allow users to freely collect images of their lives in a way that preserves privacy; both their own and the privacy of people around them. A specific and interesting problem arises when  considering the amount of time that we spend sitting in front of computer screens. In this paper we presented a potential solution to help users manage images collected from wearable devices. We show that policies based on the presence of computer screens in images can accurately be enforced at a coarse level. While fine-grained policies that work on types of screen content are more challenging to enforce, we remain optimistic based on our initial results, especially when characterizing `sensitive' vs. `non-sensitive' applications.

Much work remains to be done in this area. We are the first that we know of to apply deep learning to lifelogs in order to gain semantic meaning and offer increased levels of management and organization. Our hope is to continue this work leading to deployable systems that increase our comfort with the wearable cameras that increasingly surround us.